# TRAVEL TIME RELIABILITY IN TRANSPORTATION NETWORKS:

# A REVIEW OF METHODOLOGICAL DEVELOPMENTS

Zhaoqi Zang [a,b], Xiangdong Xu [b,*], Kai Qu [b], Ruiya Chen [b], Anthony Chen [c]

[a] School of Civil and Environmental Engineering, Nanyang Technological University, Singapore

[b] College of Transportation Engineering, Tongji University, Shanghai, China

[c] Department of Civil and Environmental Engineering, The Hong Kong Polytechnic University, Hong Kong, China

## ABSTRACT

The unavoidable travel time variability in transportation networks, resulted from the widespread supply-side and demand-side uncertainties, makes travel time reliability (TTR) be a common and core interest of all the stakeholders in transportation systems, including planners, travelers, service providers, and managers. This common and core interest stimulates extensive studies on modeling TTR. Researchers have developed a range of theories and models of TTR, many of which have been incorporated into transportation models, transport policies, and project appraisals. Adopting the network perspective, this paper aims to provide an integrated framework for summarizing the methodological developments of modeling TTR in transportation networks, including its characterization, evaluation and valuation, and traffic assignment. Specifically, the TTR characterization provides a whole picture of travel time distribution in transportation networks; TTR evaluation and TTR valuation (also known as the value of reliability, VOR) interpret abstract characterized TTR in a simple and intuitive way in order to be well understood by different stakeholders of transportation systems; and lastly TTR-based traffic assignment investigates the effects of TTR on the individual users' travel behavior and consequently the collective network flow pattern. As the above three topics are mainly separately studied in different disciplines and research areas, the integrated framework allows to better understand their relationships and may contribute to developing more possible combinations of TTR modeling philosophy. Also, the network perspective enables to pay more attention to some common challenges of modeling TTR in transportation networks, especially the uncertainty propagation from the uncertainty sources to the TTR at various spatial levels including link, route, and the entire network. Some potential directions for future research are discussed in the era of new data environment, applications, and emerging technologies.

---

[*] Corresponding author. E-mail addresses: zhaoqi.zang@ntu.edu.sg (Z. Zang), xiangdongxu@tongji.edu.cn (X. Xu), qukai@tongji.edu.cn (K. Qu), chenruiya@tongji.edu.cn (R. Chen), anthony.chen@polyu.edu.hk (A. Chen).

## 1. INTRODUCTION

As a critical factor in the efficiency and service quality of transportation networks, travel time reliability (TTR) exerts a strong influence on the stakeholders in transportation networks, including users (travelers), service providers, planners, and managers. To be specific, empirical studies have increasingly concluded that the importance of TTR is equal to or even greater than that of travel time in travelers' choice behaviors (e.g., Bates *et al*., 2001; Lam and Small, 2001; Small *et al*., 2005; Hollander, 2006; Asensio and Matas, 2008; Li *et al*., 2010, Carrion and Levinson, 2013). To account for TTR, different reliability costs may be added to different projects, and these may significantly affect the results of planers' project appraisals. Therefore, it is unsurprising that the cost of TTR is recommended or even required to be included in transportation project appraisals (de Jong and Bliemer, 2015; New Zealand Transport Agency, 2016; Organization for Economic Co-operation and Development (OECD), 2016). As for service providers and traffic managers, the reliability of travel time is one of the key performance indicators they used for monitoring or improving the service quality of transportation networks (Lyman and Bertin, 2008; Kim, 2014) and it is also their responsibility or objectives to improve the reliability of transportation systems. This common interest inspires and stimulates extensive studies on modeling TTR since Herman and Lam (1974) and Sterman and Schofer (1976) firstly recognize the need of modeling the TTR, making TTR be an active research area. We searched Scopus database for published papers using the phrase "travel time reliability" in the search category "Article title, Abstract, Keywords". There are 3,121 records, and 66.26% are published in the past ten years (i.e., 2012-2021). Figure 1 presents the total publication numbers per year from 2012 to 2021 and the top 10 transportation journals contributing to TTR, including Transportation Research Part A/B/C, Transportation, and IEEE Transactions on Intelligent Transportation Systems, etc. These results clearly show that TTR has already been a hot research topic for decades. In fact, researchers have developed a range of ideas, theories, and models of TTR, many of which have been incorporated into transportation models, transport policies, and project appraisals.

This paper aims to summarize the literature and provide a *methodological* review of modeling TTR from the *network* perspective[1]. First, the *methodological* review makes this paper concentrate on summarizing the methods of modeling TTR, especially the modeling rationales, modeling frameworks, and modeling techniques. Second, the network perspective makes this paper focus on the big picture of modeling TTR in transportation networks (with a large number

---

[1] Other definitions of reliability in networks include connectivity or terminal reliability (Wakabayashi and Iida, 1992; Bell and Iida, 1997), capacity reliability (Chen *et al*. 2002b), and travel demand satisfaction reliability (Heydecker *et al*. 2007), etc.

of diverse links, routes, and travelers), including what is TTR, how to assess TTR, and what is the behavioral and network effect of TTR. To this end, we provide an integrated framework as shown in Figure 2 to review the methodological developments in the literature from the network perspective. Specifically,

- The travel time distribution (TTD) characterization answers what is TTR through accurately capturing the manifestation of TTR. It uses a mathematical way to provide a whole picture of TTR in transportation networks, which lays the foundation of assessing TTR for different stakeholders.
- Assessing TTR can in turn interpret the abstract characterized TTD in a simple and intuitive way in order to be well understood by the different stakeholders of transportation networks. There are two ways to assess TTR, including TTR evaluation and TTR valuation, which have different applications. The former quantifies the reliability performance using various reliability measures, whereas the latter quantifies the VOR in monetary unit to understand users' behavioral response to TTR.
- The TTR has a direct and indirect effect on the travelers of a transportation network—the individual users' travel behavior and the collective network flow pattern, which corresponds to the TTR-based traffic assignment problem.

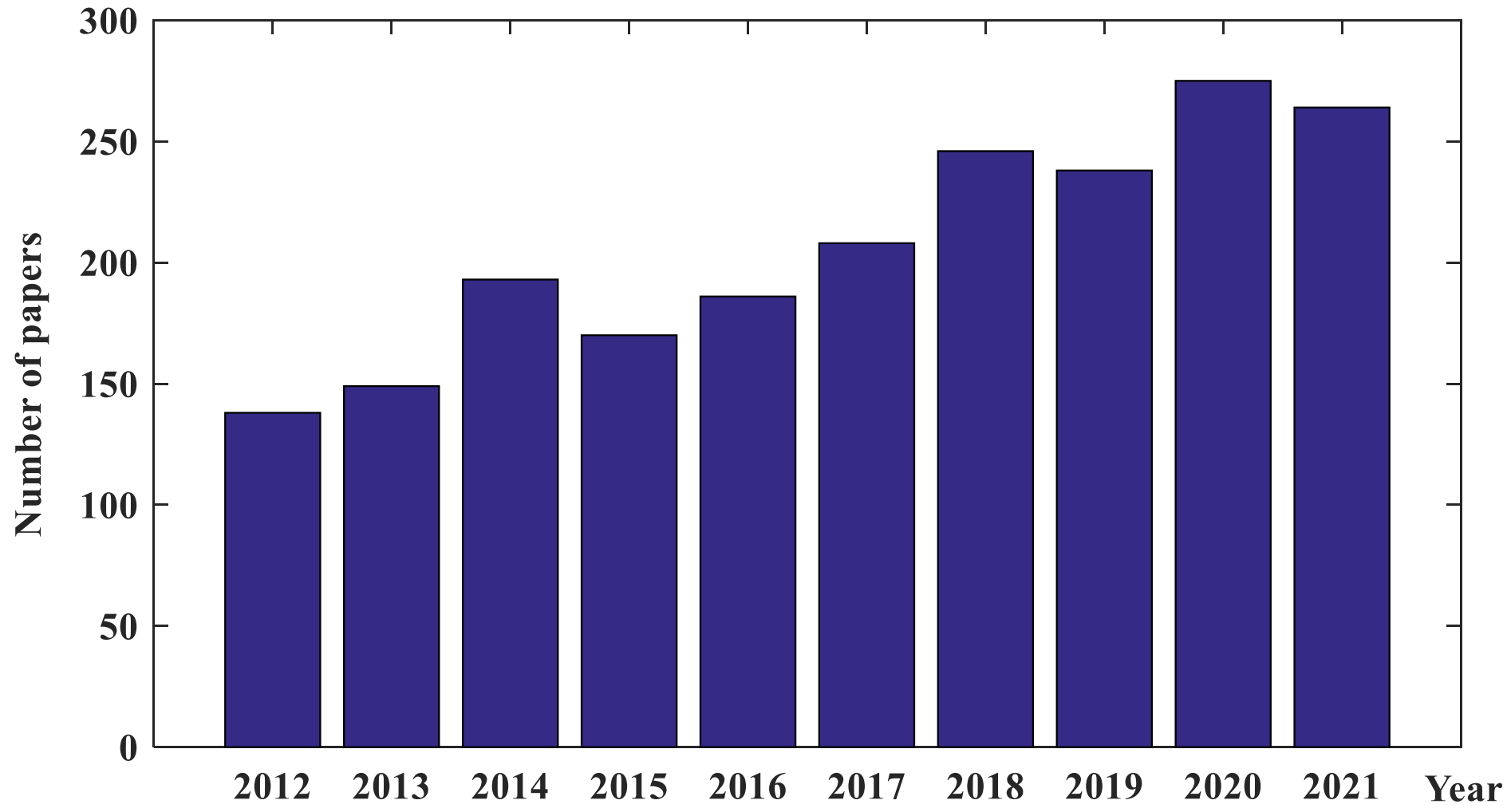

(a) Number of papers per year about travel time reliability since 2012

(b) Top 10 transportation journals contributing to travel time reliability

| Transp. Res. Rec. | Transp. Res. Part C | Transp. Res. Part B | J. Adv. Transp. | Transp. Res. Part A |
|---|---|---|---|---|
| 316 | 85 | 67 | 46 | 41 |
| J. Intell. Transp. Syst. | J. Transp. Syst. Eng. Inf. Technol. | Transportation | IEEE Trans. Intell. Transp. Syst. | J. Transp. Eng. A: Syst. |
| 40 | 36 | 33 | 31 | 29 |

Figure 1. A summary of the results of searching Scopus database for published papers using "*travel time reliability*" (Journal names are abbreviated under ISO 4 standard).

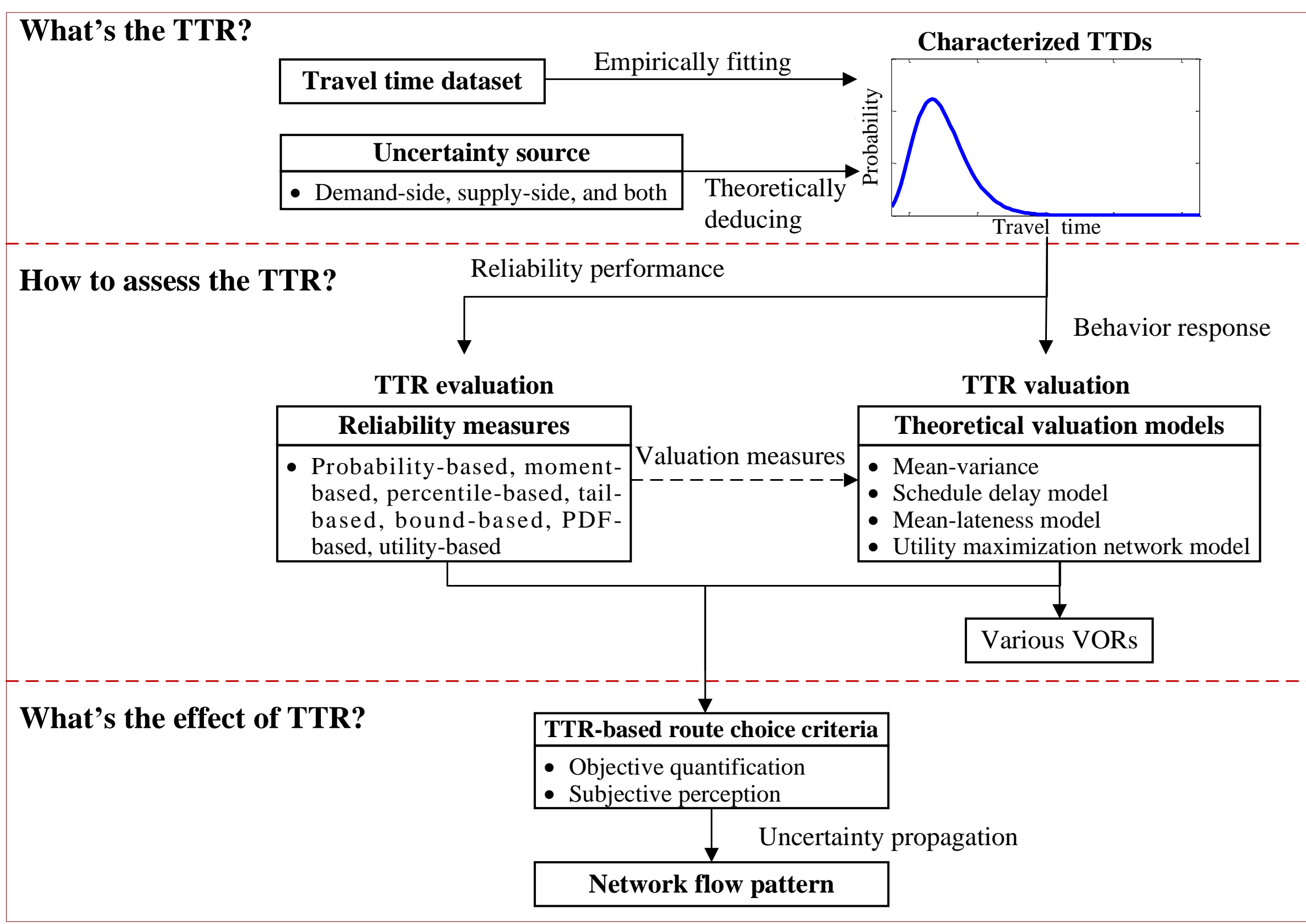

Figure 2. An integrated framework for modeling TTR in transportation networks

Although there are already some review papers about TTR, this paper contributes to the literature in several ways. First, the network perspective makes this paper put the above three research questions and topics together, which are mainly separately studied and reviewed by researchers from different disciplines and research areas, e.g., statistics, economics, and network optimization. For example, many previous studies or review papers focus on TTR valuation (e.g., Noland and Polak, 2002; Li *et al*., 2010; Small, 2012; Taylor, 2013; Wardman and Batley, 2014; Carrion and Levinson, 2012; Shams *et al*., 2017), while some papers are related to TTR evaluation (e.g., Iida, 1999; Pu, 2011; Gu *et al*., 2020) or TTR-based traffic assignment (Chen *et al*., 2011a; Nikolova and Stier-Moses, 2014). In In contrast, this paper with an integrated review framework as shown in Figure 2 allows us to better understand their relationships and to develop more possible combinations of modeling philosophy. Second, we endeavor to summarize methodological developments (e.g., modeling rationales, modeling frameworks and modeling techniques) and to provide corresponding general formulas for modeling TTR. However, previous review papers lack introduction about the modeling methods. For example, the early review on TTR by Iida (1999) only summarizes the basic concepts of TTR proposed by Asakura and Kashiwadani (1991) and Asakura (1996), while Taylor (2013) seeks to provide an overview of TTR research from travel behavior and network performance appraisal without modeling formulations. Third, the network perspective allows this paper to pay attention to the research progress about some challenges of modeling TTR in transportation networks, while previous review papers usually ignore such progress. For example, (1) the TTDs of different links and/or routes at different time periods in a transportation network are heterogeneous, and thus, the methods for characterizing TTDs in networks must be flexible enough to capture this heterogeneity while guaranteeing modeling accuracy; (2) different stakeholders in a transportation system, e.g., travelers with different preferences, service providers, planners and managers, have different understanding about the impacts of TTR on their behavioral responses, investments and decision-makings; and (3) a transportation network has a large number of links, routes, and origin-destination (O-D) pairs, which requires the modeling of TTR to explicitly capture the uncertainty propagation from the sources to the TTR at various spatial levels.

The remainder of this paper is organized as follows. Section 2 reviews the characterization of TTDs, including modeling frameworks, existing modeling rationales, and TTD models. The methods and measures for TTR evaluation are summarized in Section 3. Section 4 reviews the mathematical models, measures, and dimensions of TTR valuation. Section 5 reviews the methods used in TTR-based traffic assignment, including route choice criteria, mathematical models, and solution algorithms. Section 6 focuses on the methods for modeling uncertainty propagation, and Section 7 concludes the paper with a discussion of potential future research.

## 2. Travel Time Distributions Characterization

In this section, we discuss the meaning and necessity of characterizing TTDs, summarize the current modeling frameworks, and review the modeling rationales and corresponding TTD models[2], followed by some discussions.

Although using empirical distributions (e.g., empirical cumulative distribution function (CDF)) seems to be the most straightforward method, many studies focus on developing fitted or deduced TTD models for several reasons. First, a well-fitted TTD model and its analytical expression are key preliminaries for evaluating TTR, valuing TTR, and incorporating TTR into transportation models in transportation networks. Second, the statistical representativeness of empirical distributions cannot be guaranteed, as empirical datasets may not collect all values, especially "extreme" values that indeed existed in the tail of TTD (Taylor, 2017). Nevertheless, these "extreme" values are critical for the reliability/risk analysis of transportation networks (Xu *et al*., 2014; Taylor, 2017; Zang *et al*., 2018a; Esfeh *et al*., 2020). Last, route-level or network-level travel time datasets are not readily available for large-scale urban transportation networks, which makes it hard to directly obtain the route or network TTD models. To derive route or network TTD models, current typical method is to aggregate the TTD from link to route (or network) levels, and thus a well-fitted link TTD model is a prerequisite for this process.

[2] The TTD is also a crucial input for other calculations such as the estimation of arrival time, travel time prediction, vehicle routing problem, etc. Although all of the methods for characterizing the TTD can theoretically be used for assessing the TTR, this paper reviews only the methods under the context of TTR in transportation networks.

There are two modeling approaches for characterizing a TTD: fitting and deducing. Fitting methods use assumed distribution functions to fit travel time datasets, whereas deducing methods derive the distribution functions of travel time from assumed uncertainty sources of TTR. Hence, the inputs for models of characterizing TTD are empirical datasets or the assumed sources of variability in the fitted and deductive approaches, respectively; and the outputs are the fitted or deduced distribution functions of travel time, respectively. The distribution functions by these models include the probability density function (PDF), CDF, and inverse CDF, which is equivalent to the percentile point function (PPF).

## 2.1 Modeling Frameworks and Challenges

Figure 3 summarizes the general frameworks of the two modeling approaches to characterizing TTD in transportation networks, namely fitting and deducing. Based on Figure 3, the frameworks are described in detail below.

- For the fitting approach, the typical process for fitting TTD in transportation networks is to fit the link TTD and then to derive the route/network TTD model by aggregating the fitted link TTD models, which is marked by the blue solid line in Figure 3. The first step of this typical process is to select or develop an appropriate TTD model. Such selection depends on many factors, such as modeling rationale, the characteristics of the dataset, the application purposes, etc. Besides, it is hard to directly fit the route/network TTD due to the lack of route/network travel time datasets, although in theory, the fitted route/network TTD can be obtained from empirical route/network datasets.
- For the deducing approach, the distribution functions of the link TTD models are firstly deduced based on the assumed uncertainty sources, and then the deduced link TTD models are aggregated to derive the route/network TTD model. Section 2.2.4 will review the sources of uncertainty considered in this process and the corresponding methods for deducing the distribution functions.

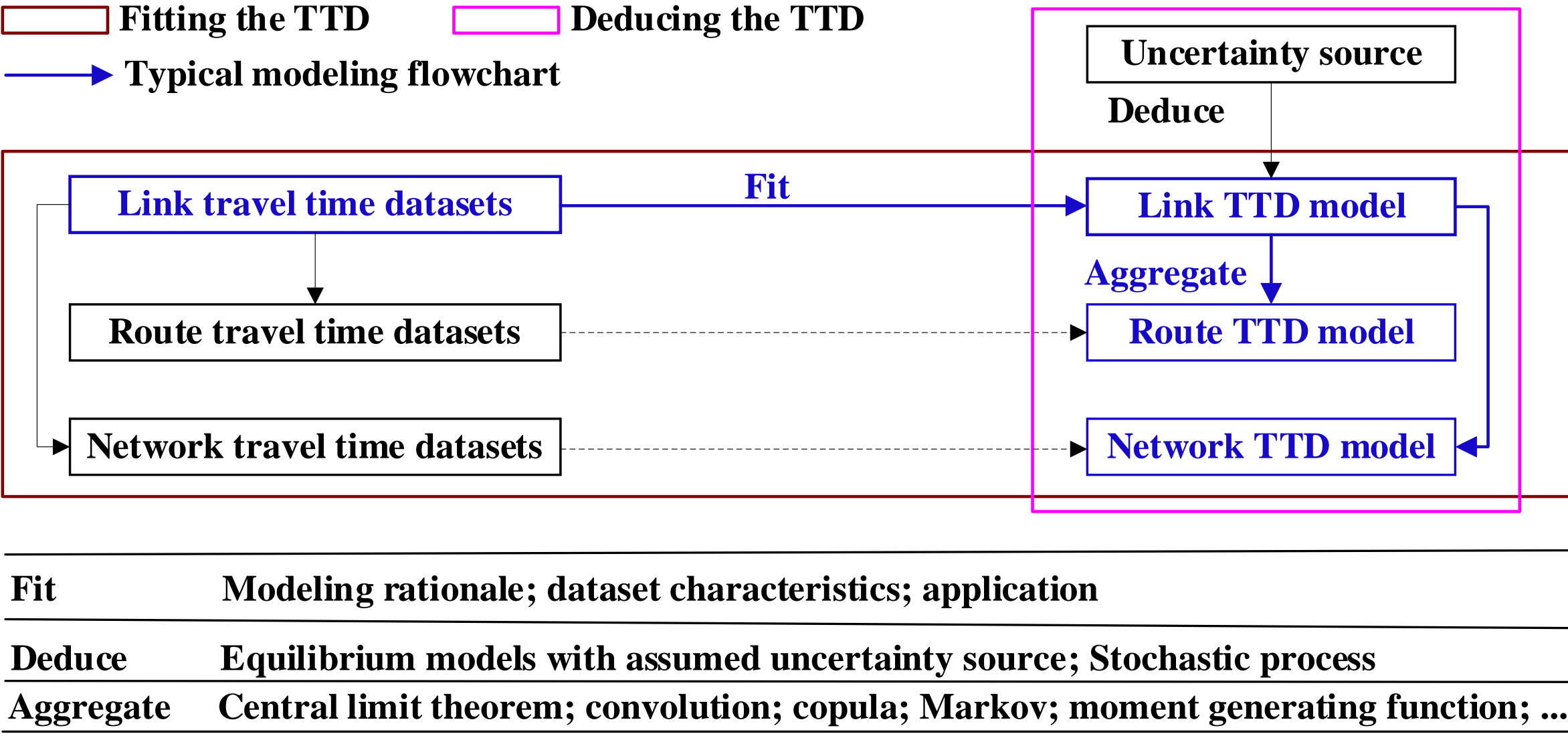

| | |
|---|---|
| **Fit** | **Modeling rationale; dataset characteristics; application** |
| **Deduce** | **Equilibrium models with assumed uncertainty source; Stochastic process** |
| **Aggregate** | **Central limit theorem; convolution; copula; Markov; moment generating function; ...** |

Figure 3. Modeling frameworks for characterizing travel time distributions (TTDs)

Some challenges exist in characterizing TTDs under these two frameworks as follows.

- Because of their heterogeneity, TTDs can have various statistical features, i.e., symmetric or asymmetric, left- or right-skewed, long or short tail, and unimodal or multimodal, which makes characterizing TTDs very challenging. Many empirical studies have verified that TTDs are right-skewed with a long/fat tail (Polus, 1979; Fosgerau and Fukuda, 2012; Susilawati *et al*., 2013; Srinivasan *et al*., 2014; Kim and Mahmassani, 2015; Delhomme *et al*., 2015; Taylor, 2017). However, despite with this typical feature, some TTDs may present quite different statistical features, e.g., TTDs may be left-skewed or symmetric (van Lint and van Zuylen, 2005; van Lint *et al*., 2008; Yang *et al*., 2014b; Zang *et al*., 2018a). van Lint and van Zuylen (2005) and Zang *et al*. (2018a) identify TTDs with negative skewness, near-zero skewness, positive skewness, and strong positive skewness. Other studies further identify the multimodality of TTDs in urban transportation systems (e.g., Dong and Mahmassani, 2009; Guo *et al*., 2010; Taylor and Somenahalli, 2010; Kazagli and Koutsopoulos, 2012; Chen *et al*., 2014b; Yang *et al*., 2014a).
- The theoretical and empirical applications of TTD models have different requirements. The TTD models used in empirical applications focus on accuracy and computational efficiency, which are non-trivial problems because of the heterogeneity of TTDs. In contrast, theoretical applications use closed-form expressions of TTD models as the basis for conceptualizing and formulating computationally tractable or analytically derived traffic models that incorporate TTR.
- Both the fitting and deducing approaches derive a route/network TTD model from the link TTD model, which involves aggregating uncertainty from the link level to the route/network level. As transportation networks often consist of a large number of links, modeling such uncertainty aggregation is quite challenging, which requires a trade-off between computational efficiency and modeling accuracy. It should be noted that uncertainty aggregation is also an important part of TTR evaluation and TTR-based traffic assignment models.

Table 1 lists the models for characterizing TTDs in transportation networks developed within these two modeling frameworks, indicating their modeling rationales, applications, closed-form formulae, and forms of random variables. Section 2.2 summarizes the four modeling rationales underlying these two modeling frameworks, and presents a general formula for each modeling rationale and TTD models that correspond to each modeling rationale. Section 2.3 discusses how studies overcome the first two challenges in characterizing TTD outlined above, and the methods for addressing the third challenge are reviewed in Section 6.

## 2.2 Modeling Rationales and Corresponding TTD Models

The four modeling rationales underlying the models for characterizing TTD are the single distribution modeling rationale, the mixture distribution modeling rationale, the moment-based modeling rationale, and the source-based derivation modeling rationale. The first three are the basis for models that fit TTD based on travel time datasets, and the last is the basis for models that deduce TTD from an assumed uncertainty source.

Let a unified notation, i.e., $z$, denote the general formulation for each of the four modeling rationales. Assume that $T$ is a random variable that denotes travel time or its variants, $\rho$ denotes the probability or confidence level, and let $f(T)$, $F(T)$, and $F^{-1}(\rho)$ denote the PDF, CDF, and PPF, respectively. The variants of travel time in TTD models include standardized travel time, total travel time, pace, and (extreme) travel time delay. Standardized travel time can be obtained by (Travel time – mean)/standard deviation and is usually used in VOR research (Fosgerau and Fukuda, 2012; Taylor, 2017; Zang *et al*., 2018a, 2018b; Li, 2019), which corresponds to the general assumption discussed in TTR valuation, i.e., the standardized travel time distribution is independent of departure time. Total travel time is the sum of all users' travel times in a transportation network, and thus it is used in network-wide reliability models (Chen *et al*., 2014a; Xu *et al*., 2013, 2014). Pace equals the random travel time divided by link/path length and is used to exclude the travel time variation that arises from variations in link length (Daganzo, 1997; Mahmassani *et al*., 2013; Saberi *et al*., 2014). The difference between (extreme) travel time and the minimum (extreme) travel time is (extreme) travel time delay, which is used by Kim and Mahmassani (2015) and Esfeh *et al*. (2020) to explore the direct proportionality between the mean and standard deviation of (extreme) travel time.

### 2.2.1 Single distribution modeling rationale

The classic single distribution model is the most widely used method to characterize TTR. Its general formula can be written as

$$z = f(T \mid \theta, l, k) \tag{1}$$

where $\theta$ is the scaling factor, $l$ is the location factor, and $k$ is the shape factor. Under this modeling rationale, two-parameter statistical distribution functions such as the Normal (Bell and Iida, 1997), Lognormal (Emam and Al-Deek, 2006), Weibull (Al-Deek and Emam, 2006), and Gamma distributions (Polus, 1979) are first used to fit the TTDs[3]. However, the verification of the heterogeneity of TTDs by many empirical studies has led to the adoption of distribution functions with three or more parameters to accurately capture the TTDs' multiple statistical properties, e.g., shifted Lognormal (Srinivasan *et al*., 2014), compound Gamma (Kim and Mahmassani, 2015), Generalized Beta (Castillo *et al*., 2012), Stable (Fosgerau and Fukuda, 2012), (compound) generalized extreme value (Lei *et al*., 2014), Burr (Susilawati *et al*., 2013)[4], and tailored Wakeby-type distribution (Zhang *et al*., 2018). Note that compound distribution is the probability distribution with the assumption that its random variable follows a distribution type with an unknown parameter that also follows the same distribution type. Therefore, the compound Gamma distribution and compound generalized extreme value are still viewed as the models under the single distribution modeling rationale.

### 2.2.2 Mixture distribution modeling rationale

It is difficult for single distribution models to characterize the multimodal travel time resulting from interrupted flows at intersections or congested traffic conditions in urban transportation systems (e.g., Dong and Mahmassani, 2009; Guo *et al*., 2010; Taylor and Somenahalli, 2010; Kazagli and Koutsopoulos, 2012; Chen *et al*., 2014b; Yang *et al*., 2014a). However, mixture distribution models can deal with such multimodality of TTDs. The mixture model is usually a weighted combination of serval single distribution models of the same type and the component distribution can be any of the unimodal TTD models developed under the single distribution modeling rationale. Its general formula is

$$z = \sum \lambda_i f_i(T \mid \theta_i, l_i, k_i) \tag{2}$$

where $\lambda_i$ is the $i$th weight associated with the $i$th single distribution. Mixture distribution models can establish a connection between TTDs and the related traffic states, which leads to a better fitting for multimodal TTDs (Guo *et al*., 2010; Rahmani *et al*., 2015; Chen *et al*., 2017). The models based on this modeling rationale include the Normal mixture model (Guo *et al*., 2010), Lognormal mixture model (Kazagli and Koutsopoulos, 2012), Gamma mixture model (Yang and Wu, 2016), finite mixture of the regression model (Chen *et al*., 2014b), and kernel density estimation (Fosgerau and Fukuda, 2012; Li, 2019)[5].

---

[3] For each TTD model, we list only one or two representative references in the text. Table 1 lists more related papers.

[4] The Burr distribution is also referred to as the Singh–Maddala distribution. The latter is used by Guessous *et al*. (2014) to characterize TTD.

### 2.2.3 Moment-based modeling rationale

Both the single distribution and mixture distribution modeling rationales need a prior assumption about the distribution type that travel time follows, which is unknown in reality. Furthermore, empirical studies show that although TTDs are typically right-skewed with long/fat tails, they can also be left-skewed or have near-normal distributions (van Lint and van Zuylen, 2005; van Lint *et al*., 2008; Yang *et al*., 2014b; Zang *et al*., 2018a). To circumvent these challenges, some studies directly use statistical information of travel time datasets to fit heterogeneous TTDs, avoiding the need to assume the distribution type. Moments are the most common statistical information used under this modeling rationale, and thus we refer to this as the moment-based modeling rationale. The general formula, which is based on the first $n$ moments, can be expressed as

$$z = f(T \mid \xi_1, \xi_2, \ldots, \xi_n) \tag{3}$$

where $\xi_i$ is the $i$th moment of travel time. The Cornish–Fisher expansion (Zang *et al*., 2018a), Gram–Charlier expansion (Hou and Tan, 2009), and Johnson curves (Clark and Watling, 2005) are three representative methods, and they all use the first four moments of travel time, i.e., mean, standard deviation, skewness, and (excess) kurtosis. Ng *et al*. (2011) uses the probability inequalities based on the first $n$ moments to obtain the upper bounds of the tail probabilities instead of an exact probability distribution.

### 2.2.4 Source-based derivation modeling rationale

The above three modeling rationales use empirical datasets to fit TTDs. An alternative rationale is the source-based derivation modeling rationale that uses assumptions about the uncertainty sources of TTR to deduce TTD models. The uncertainty sources of demand and/ or supply are main assumptions, e.g., the link capacity variations belonging to supply-side uncertainty in Lo and Tung (2003) and Lo *et al*. (2006), the day-to-day demand fluctuations belonging to demand-side uncertainty in Clark and Watling (2005) and Shao *et al*. (2006a, 2006b), and the

---

[5] The kernel density estimation is also a combination of its component kernels, so it is considered as a method of the mixture distribution modeling rationale despite that the component kernel may not be a distribution function.

Table 1. A summary of existing TTD models for TTR in transportation networks

| TTD model | Modeling Rationale | Formula (PDF) | Random variable | Application | Closed-forms | References |
|---|---|---|---|---|---|---|
| Normal | Single distribution | $f(T\,|\theta, l)$ | TT | Theoretical | PDF<br>CDF<br>PPF | Bell and Iida (1997); Lam and Xu (1999); Yin and Ieda (2001); Lomax *et al*. (2003); Lo and Tung (2003); Watling (2006); Siu and Lo (2013) |
| Lognormal | Single distribution | $f(T\,|\theta, k)$ | (T)TT<br>TT | Theoretical<br>Empirical | PDF<br>CDF | Kaparias *et al*. (2008); Pu (2011); Chen *et al*. (2014b)<br>Herman and Lam (1974); Richardson and Taylor (1978); Emam and Al-Deek (2006); Rakha *et al*. (2006, 2010); van Lint *et al*. (2008); Arezoumandi (2011) |
| Shifted Lognormal | Single distribution | $f(T\,|\theta, l, k)$ | TT | Empirical<br>Theoretical | PDF<br>CDF | Srinivasan *et al*. (2014)<br>Lee *et al*. (2019) |
| Weibull | Single distribution | $f(T\,|\theta, k)$ | TT | Empirical | PDF<br>CDF<br>PPF | Al-Deek and Emam (2006) |
| Gamma | Single distribution | $f(T\,|\theta, k)$ | TT | Empirical | PDF<br>CDF | Polus (1979) |
| Compound Gamma | Single distribution | $f(T\,|\theta_1, \theta_2, k)$ | TD/dist. | Empirical | None | Kim and Mahmassani (2015) |
| Generalized Beta | Single distribution | $f(T\,|\theta, l, k)$ | TT | Theoretical<br>Empirical | PDF | Castillo *et al*. (2012) |
| Stable | Single distribution | $f(T\,|\theta, l, k_1, k_2)$ | STT | Empirical | None | Fosgerau and Fukuda (2012) |
| Burr | Single distribution | $f(T\,|\theta, k_1, k_2)$ | (S)TT | Empirical | PDF<br>CDF<br>PPF | Taylor (2012, 2017); Susilawati *et al*. (2013); Guessous *et al*. (2014) |
| Tailored Wakeby-type | Single distribution | $f(T\,|\theta, l_1, l_2, k_1, k_2)$ | Bus<br>HAR | Empirical | PPF | Zhang *et al*. (2018) |

| | | | | | | |
|---|---|---|---|---|---|---|
| Generalized extreme value (GEV) | Single distribution | $f(T \mid \theta, l, k)$ | Pace | Empirical | PDF<br>CDF<br>PPF | Lei *et al*. (2014); Zhang *et al*. (2019b) |
| Generalized Pareto | Single distribution | $f(T \mid \theta, l, k)$ | Pace | Empirical | PDF<br>CDF<br>PPF | Lei *et al*. (2014) |
| Compound GEV | Single distribution | $f(T \mid l_1, l_2, k_1, k_2, s_\mu)$ | ETD | Empirical | None | Esfeh *et al*. (2020) |
| Mixture model | Mixture distribution | $\sum_i \lambda_i f(T \mid \theta_i, l_i, k_i)$ | TT | Empirical | PDF | Dong and Mahmassani (2009); Jintanakul *et al*. (2009); Guo *et al*. (2010); Taylor and Somenahalli (2010); Guo *et al*. (2012); Kazagli and Koutsopoulos (2012); Chen *et al*. (2014b); Yang *et al*. (2014a); Ma *et al*. (2016); Yang and Wu (2016) |
| Kernel density estimation | | | (S)TT | | | Fosgerau and Fukuda (2012); Yang *et al*. (2014b); Rahmani *et al*. (2015); Li (2019) |
| Gram–Charlier expansion | Moment-based | $f(T \mid \xi_1, \xi_2, \xi_3, \xi_4)$ | TT | Empirical | PDF | Hou and Tan (2009) |
| Johnson curves | Moment-based | $f(T \mid \xi_1, \xi_2, \xi_3, \xi_4)$ | TTT | Theoretical | PDF<br>CDF<br>PPF | Clark and Watling (2005) |
| Cornish–Fisher expansion | Moment-based | $f(T \mid \xi_1, \xi_2, \xi_3, \xi_4)$ | (T)TT | Theoretical | PPF | Lu *et al*. (2005, 2006); Di *et al*. (2008); Chen *et al*. (2011b); Xu *et al*. (2013, 2014) |
| | | | (S)TT | Empirical | | Zang *et al*. (2018a, 2018b) |
| N.A. | Sourced-based derivation | N.A. | TT | Empirical | PDF | Kharoufeh and Gautam (2004); Zheng *et al*. (2017) |
| | Sourced-based derivation and single distribution | | TTT | Theoretical | N.A. | Lo and Tung (2003); Lo *et al*. (2006); Shao *et al*. (2006a, 2006b); Lam *et al*. (2008); Ng and Waller (2010a); Li *et al*. (2017) |

Note: 1. $\theta$ is the scaling factor; $l$ is the location factor; $k$ is the shape factor; $s_\mu$ is the seasonal mean of the ETD in compound GEV; "None" means that it does not have any closed-form formulae; N.A. means that this column is not applicable.
2. TT: travel time; STT: standardized travel time; TTT: total travel time; (E)TD: (extreme) travel time delay; HAR: headway adherence ratio; dist.: distance.

adverse weather conditions with different rainfall intensities belonging to both demand-side and supply-side uncertainties in Lam *et al*. (2008) and Li *et al*. (2017). Interested readers may refer to van Lint *et al*. (2008) and Chen and Zhou (2010) for a more complete summary about the typical uncertainty factors of supply side or demand side causing travel time variability.

Generally, this modeling rationale is used to derive route or network TTDs for assessing the network reliability. The general formula for these models depends on the assumed uncertainty source and the assumed link travel time function. Suppose that the link travel time $T$ is a function of the link flow $v$ and link capacity $C$; then, we have the following formula for link $a$:

$$T_a = t_a(v_a, C_a),\ \forall\ a \in A \tag{4}$$

where $A$ is the set of links in a transportation network.

Different assumed sources of uncertainty lead to different expressions of the random link travel time. To illustrate this, we use a tilde to highlight the stochasticity of this variable due to the assumed source of uncertainty. Let $t_a(\tilde{v}_a, C_a)$, $t_a(v_a, \tilde{C}_a)$, and $t_a(\tilde{v}_a, \tilde{C}_a)$ represent the deduced link travel time function based on demand-side uncertainty, supply-side uncertainty, and both demand-side and supply-side uncertainties, respectively. Then, the random link travel time is expressed as

$$\tilde{T}_a = t_a(\tilde{v}_a, C_a) \text{ or } t_a(v_a, \tilde{C}_a) \text{ or } t_a(\tilde{v}_a, \tilde{C}_a),\ \forall a \in A \tag{5}$$

Taking $t_a(v_a, \tilde{C}_a)$ as an example, the mean and variance of the link travel time, i.e., $\mu_T^a$ and $(\sigma_T^a)^2$, can be computed as follows:

$$\begin{aligned} \mu_T^a &= E[\tilde{T}_a] = E[t_a(v_a, \tilde{C}_a)],\ \forall a \in A \\ (\sigma_T^a)^2 &= Var[\tilde{T}_a] = Var[t_a(v_a, \tilde{C}_a)],\ \forall a \in A \end{aligned} \tag{6}$$

Now, we have expressions of link travel time and corresponding mean travel time and travel time variance. Below, we briefly introduce the general formula of route or network TTDs.

For transportation networks with routes that consist of many links, the travel time of route $j$ between O-D pair $\omega$ can be obtained by summing the corresponding link travel time:

$$\tilde{T}_{\omega j} = \sum_{a \in A} \tilde{T}_a \delta_{aj},\ \forall j \in J_\omega,\ \forall \omega \in \Omega_\omega \tag{7}$$

where $J_\omega$ and $\Omega_\omega$ are the set of routes between O-D pair $\omega$ and the set of O-D pairs in the network, $\delta_{aj}$ is the link-route incidence indicator, and $\delta_{aj}$ = 1 if link $a$ is on route $j$, and 0 otherwise. Similarly, the route travel time is a random variable, so the mean and the variance, denoted as $\mu_T^{\omega j}$ and $\left(\sigma_T^{\omega j}\right)^2$, of the route travel time can be expressed as

$$\mu_T^{\omega j} = \sum_{a \in A} \mu_T^a \delta_{aj}, \ \forall j \in J_\omega, \ \forall \omega \in \Omega_\omega \ \text{ and}$$
$$\left(\sigma_T^{\omega j}\right)^2 = \sum_{a \in A} \delta_{aj} Var\left[\tilde{T}_a\right] = \sum_{a \in A} \delta_{aj} \left(\sigma_T^a\right)^2, \ \forall j \in J_\omega, \ \forall \omega \in \Omega_\omega \tag{8}$$

With calculated mean and variance, many methods have been developed to derive route TTDs including Central Limit Theorem, cupula, convolution, Markov chain, etc., to be reviewed in detail in Section 6.2. Taking the Central Limit Theorem as an example, the route travel time would follow a Normal distribution regardless of the link TTDs (e.g., Lo and Tung, 2003; Lo *et al*., 2006; Shao *et al*., 2006a, 2006b). Consequently, the route TTD can be expressed as

$$\tilde{T}_{\omega j} \sim N\left(\mu_T^{\omega j}, \left(\sigma_T^{\omega j}\right)^2\right), \ \forall j \in J_\omega, \ \forall \omega \in \Omega_\omega \tag{9}$$

For transportation networks consisting of many links, the total travel time is the product of the link flow and the link travel time of all of the links:

$$\tilde{T} = \sum_{a \in A} \tilde{T}_a \tilde{v}_a \tag{10}$$

Then, we can deduce the formulae of the first $n$ moments of total travel time:

$$\xi_1 = E\left[\tilde{T}\right], \ \xi_2 = E\left[\tilde{T}^2\right], \ \ldots, \ \xi_n = E\left[\tilde{T}^n\right] \tag{11}$$

With the deduced moments of TTT, the methods mentioned in Section 2.2.3 can be directly used to derive the distribution function of total travel time:

$$z = f\left(T \mid E\left[\tilde{T}\right], \ E\left[\tilde{T}^2\right], \ldots, E\left[\tilde{T}^n\right]\right) \tag{12}$$

It should be noted that in the literature, the stochastic traffic process is another typical assumption regarding the origin of travel time uncertainty at the operational level in urban transportation systems. For example, Kharoufeh and Gautam (2004) assumes that a vehicle's speed follows a stochastic speed process, and they use a partial differential equation and Laplace transforms to derive the link TTDs. Zheng and van Zuylen (2010) and Zheng *et al*. (2017) assume that urban travel times are the result of many stochastic factors (e.g., stochastic traffic flow and arrivals, and departures) and traffic control, and they use shockwave theory to derive the TTDs of urban signalized arterial roads.

Table 2. Closed-form formulations of travel time distribution (TTD) models

| Distribution | Formula (PDF or PPF) | Distribution | Formula (PDF or PPF) | Distribution | Formula (PDF or PPF) |
| --- | --- | --- | --- | --- | --- |
| Normal | $f\left(T \mid \theta, l\right)=\frac{1}{\sqrt{2\pi}\theta}e^{-\frac{1}{2}\left(\frac{T-l}{\theta}\right)^2}$ | Lognormal | $f\left(T \mid \theta, k\right)=\frac{1}{Tk\sqrt{2\pi}}e^{-\frac{1}{2}\left(\frac{\ln T-k}{\theta}\right)^2}$ | Shifted Lognormal | $f\left(T \mid \theta, l, k\right)=\frac{1}{\left(T-l\right)k\sqrt{2\pi}}e^{-\frac{1}{2}\left(\frac{\ln(T-l)-k}{\theta}\right)^2}$ |
| Weibull | $f\left(T \mid \theta, k\right)=\frac{k}{\theta}\left(\frac{T}{\theta}\right)^{k-1}e^{-\left(\frac{T}{\theta}\right)^k}$ | Gamma | $f\left(T \mid \theta, k\right)=\frac{1}{\Gamma\left(k\right)\theta^k}T^{k-1}e^{-\frac{T}{\theta}}$ | Mixture model | $\sum_i \lambda_i f_i\left(T \mid \theta_i, l_i\right)=\sum_i \lambda_i \frac{1}{\theta_i\sqrt{2\pi}}e^{-\frac{1}{2}\left(\frac{T-l_i}{\theta_i}\right)^2}$ |
| Kernel density estimation | $f\left(T \mid \theta, l, k\right)=\frac{1}{k\theta}\sum_{i=1}^{k}K\left(\frac{T-l_i}{\theta}\right)$ | Burr | $f\left(T \mid \theta, k_1, k_2\right)=\frac{k_1k_2}{\theta}\left(\frac{T}{\theta}\right)^{k_1-1}\left(1+\left(\frac{T}{\theta}\right)^{k_1}\right)^{-\left(k_2+1\right)}$ | Generalized Beta | $f\left(T \mid \theta, l, k_1, k_2\right)=\frac{\Gamma\left(k_1+k_2\right)}{\Gamma\left(k_1\right)\Gamma\left(k_2\right)}\left(\frac{T-l}{\theta}\right)^{k_1-1}\left(1-\frac{T-l}{\theta}\right)^{k_2-1}$ |
| Tailored Wakeby-type | $F^{-1}\left(\rho \mid l, \theta_1, \theta_2, k_1, k_2\right)=l+\frac{\theta_1}{k_1}\left(1-\left(1-\rho\right)^{k_1}\right)+\frac{\theta_2}{k_2}\left(1-\left(1-\rho\right)^{-k_2}\right)$ | | | | |
| Generalized Extreme Value | $f\left(T \mid \theta, l, k\right)=\frac{1}{\theta}\phi\left(T\right)^{k+1}e^{-\phi(T)}$, where $\phi\left(T\right)=\left(1+k\left(\frac{T-l}{\theta}\right)\right)^{-1/k}$ $\left(\text{if } k \neq 0\right)$ or $\phi\left(T\right)=e^{-(T-l)/\theta}$ $\left(\text{if } k=0\right)$ | | | | |
| Generalized Pareto | $f\left(T \mid \theta, l, k\right)=\frac{1}{\theta}\left(1+k\left(\frac{T-l}{\theta}\right)\right)^{-\left(k+1\right)/k}$, for $T>l$ when $k>0$ and $l \leq T \leq l-\theta/k$ when $k<0$ | | | | |
| Gram–Charlier | $f\left(T \mid \xi_1, \xi_2, \xi_3, \xi_4\right)=\frac{1}{\sqrt{2\pi}}e^{-\frac{1}{2}\phi^2}\left(1+\frac{\xi_3}{6}\left(\phi^3-3\phi\right)+\frac{\xi_4}{24}\left(\phi^4-6\phi^2+3\right)\right)$, where $\phi=\left(T-\xi_1\right)/\xi_2$ | | | | |
| Cornish–Fisher | $F^{-1}\left(\rho \mid \xi_1, \xi_2, \xi_3, \xi_4\right)=\xi_1+\xi_2\left(U_\rho+\xi_3\left(U_\rho^2-1\right)/6+\xi_4\left(U_\rho^3-3U_\rho\right)/24-\xi_3^2\left(2U_\rho^3-5U_\rho\right)/36\right)$, where $U\rho$ is $\rho$ quantile of the standard Normal. | | | | |

Note: 1. The formula of the mixture model depends on the single distribution element; here, we provide the formula of the mixture Normal model as an example.
2. The Compound Gamma distribution does have a closed-from PDF, but cannot be expressed using one formulation. Please refer to Kim and Mahmassani (2015) for details.

### 2.3 Discussions of the Existing TTD Models

#### 2.3.1 Capturing the heterogeneous TTDs based on different modeling rationales

As verified by Plötz *et al.* (2017) and Zang *et al.* (2018a), the performance of the same TTD model in fitting different travel time datasets may be different. Consequently, many different TTD models, based on different modeling rationales, have been developed to fit heterogeneous TTDs. As can be seen from Table 1, the TTD models based on single distribution and mixture modeling rationales have been widely studied for decades. In particular, the single distribution model is suitable for fitting a TTD with a typical right-skewed distribution and a long/fat tail, whereas the mixture distribution model can deal with the multimodality that arises from interrupted flow at intersections or congested traffic conditions in the urban transportation systems. Although relatively less attention has been paid to TTD models based on the moment-based modeling rationale, the performance of TTD models based on this modeling rationale are promising because they do not need assumptions about the distribution of travel time and can adaptively fit heterogeneous TTDs using the actual travel time data. The fourth set of methods to derive TTD models, based on the source-based derivation modeling rationale, is mainly used in network-wide theoretical studies.

Furthermore, as shown in Table 1, some TTD models include multiple modeling rationales. For example, the widely used Normal and Lognormal distributions used in the single distribution modeling rationale are usually combined with an assumed uncertainty source of travel time variability in TTD models (e.g., Lo and Tung, 2003; Lo *et al*., 2006; Shao *et al*., 2006a, 2006b; Lam *et al*., 2008; Li *et al*., 2017). In addition, some TTD models adopt both source-based derivation and moment-based modeling rationales. For instance, considering day-to-day demand fluctuations, Clark and Watling (2005) applies Johnson curves based on the first four moments of total travel time to characterize their TTD model for evaluating network-wide TTR.

#### 2.3.2 Requirements of TTD models for different applications

TTD models have both theoretical and empirical applications. In theoretical applications, researchers directly assume the distribution type of the travel time and use this to explore further theoretical components of TTR-based models such as travel behavior (Siu and Lo, 2013), reliable route finding (Chen and Ji, 2005; Chen *et al*., 2014a; Srinivasan *et al*., 2014; Lee *et al*., 2019), and network equilibrium (Watling, 2006; Castillo *et al*., 2012). In empirical applications, researchers develop TTD models to fit real travel time datasets (e.g., Fosgerau and Fukuda, 2012; Susilawati *et al*., 2013; Zang *et al*., 2018a, 2018b). Obviously, the criteria for TTD models for these two applications are different. Desirable mathematical properties are crucial for TTD models used in theoretical applications, whereas TTD models for empirical applications must be accurate and computationally efficient. Therefore, the Normal and Lognormal distributions are the two most widely used TTD models in theoretical applications,[6] but more advanced or complicated TTD models based on the mixture or moment-based methods are more common in empirical applications and seldom used in theoretical applications, with an exception of the Cornish–Fisher expansion (e.g., Lu *et al*., 2005, 2006; Di *et al*., 2008, Chen *et al*., 2011b, Xu *et al*., 2013, 2014).

The closed-form expressions of TTD models are critical for assessing TTR or including TTR into traffic models. For example, as Zang *et al*. (2018a) notes (and as to be shown in Table 3), the PPF (i.e., travel time percentile function) is a basic and common element of most TTR measures. Because of the existence of the closed-form PPF, the Burr distribution and Cornish–Fisher expansion are preferred for calculating Fosgearu's TTR ratio (Taylor, 2017; Zang *et al*., 2018a). In addition, the closed-form expression of the TTD model is the basis for conceptualizing and formulating computationally tractable or analytically derived traffic models that consider TTR. Thus, the Normal distribution has been widely used in TTR-based traffic models due to its simple and closed-form PDF (e.g., Lam and Xu, 1999; Yin and Ieda, 2001; Lo and Tung, 2003; Lo *et al*., 2006; Watling, 2006; Lam *et al*., 2008; Siu and Lo, 2013), although very few studies use it to empirically fit TTD in the context of TTR.

Table 1 documents whether existing TTD models have closed-form expressions of PDF, CDF, and PPF. For reference, Table 2 presents the models with closed-form expressions of the PDF or the inverse CDF. As shown in Table 1, the modeling rationale and closed-form formula of TTD models are closely related. Specifically, most TTD models based on the single distribution modeling rationale have at least one closed-form expression, e.g., Normal, Lognormal, shifted Lognormal, Weibull, Gamma, Generalized Beta, Burr, tailored Wakeby-type, and Generalized extreme value. However, the Stable, Compound Gamma and Compound Generalized extreme value distributions do not have any closed-form expressions, although they are based on the single distribution modeling rationale. Most of the models that adopt the mixture distribution modeling rationale have only closed-form PDF. Similarly, models based on moment-based methods have only closed-form PPFs. As for TTD models based on the source derivation modeling rationale, Kharoufeh and Gautam (2004) derives a closed-form expression of CDF, whereas Zheng *et al*. (2017) derives a closed-form expression of PDF. For TTD models based on both single distribution and source derivation modeling rationales, the existence of closed-form expressions depends on the assumed distribution type of the unimodal TTD model. For example, the TTD models in Lo and Tung (2003), Lo *et al*. (2006), Shao *et al*. (2006a, 2006b), and Lam *et al*. (2008) have a closed-form PDF due to the assumed Normal distribution, and the TTD model in Li *et al*. (2017) has a closed-form PDF due to the assumed Lognormal distribution.

[6] The earliest adoption of Normal distribution in travel time reliability can only date back to Bell and Iida (1997) in which Normal distribution was introduced to characterize TTD model.

## 3. Travel Time Reliability Evaluation

In this section, we review the modeling perspectives and the corresponding methods used in TTR evaluation, and then summarize and analyze the reliability measures.

As characterized TTD is the basis of TTR evaluation in transportation networks, the two common modeling perspectives of evaluating TTR are each associated with one of the modeling perspectives of characterizing TTD: (1) evaluating TTR directly using travel time datasets, and (2) evaluating TTR based on assumed distributions of the sources of uncertainty. Figure 4 summarizes these two modeling perspectives, giving the methods used in each for developing reliability measures to evaluate TTR. Specifically, directly using travel time data is preferred when appropriate datasets are available, as this can eliminate the need to assume the probability distributions of the sources of uncertainty and automatically account for all sources of uncertainty. Evaluations based on assumptions about the sources of uncertainty are useful alternatives, especially for large-scale networks without sufficient travel time data. However, the accuracy of the assumed distributions is not guaranteed when there is no (or insufficient) data available for calibration.

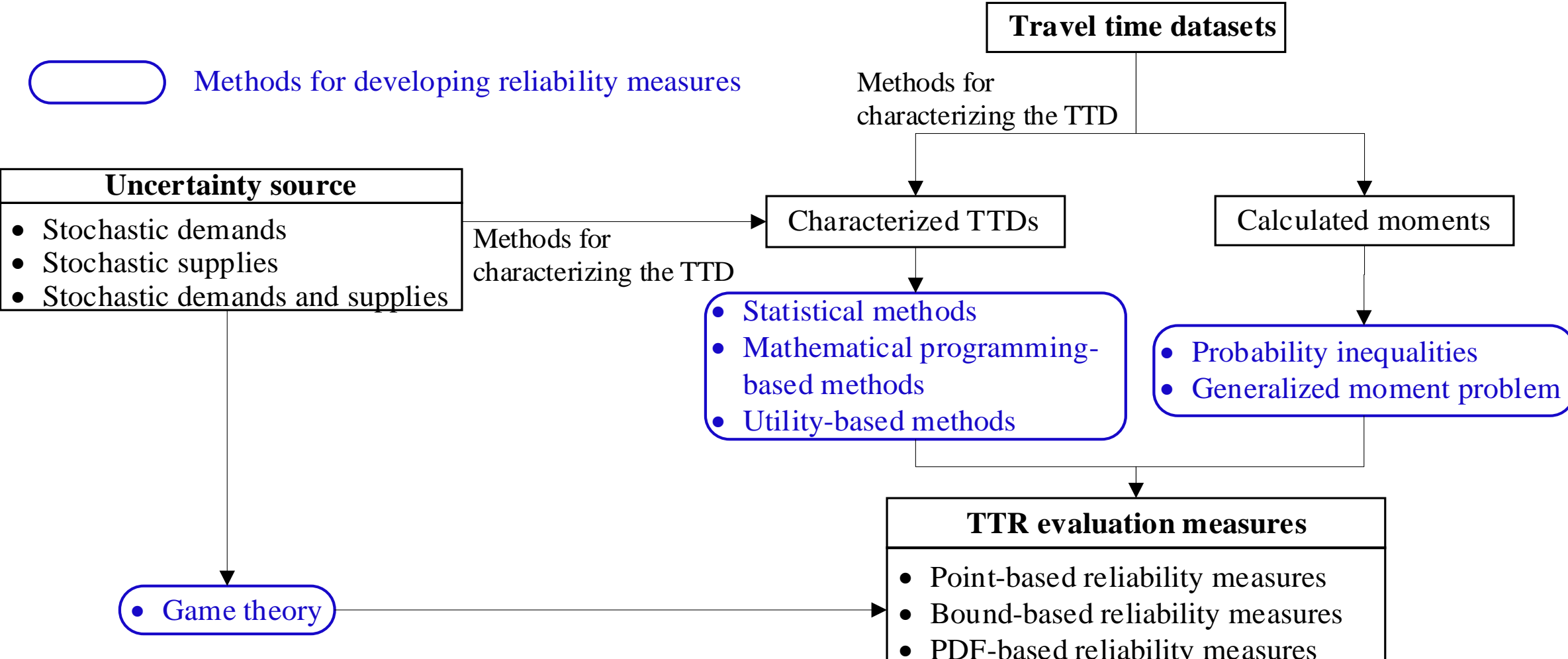

Figure 4. Frameworks and methods for evaluating TTR

Below is a detailed discussion of the methods used in these two modeling perspectives for TTR evaluation.

- There are two ways to develop reliability measures to evaluate TTR directly using travel time data. The first is to directly develop reliability measures using the moment information of the travel time datasets. Current methods include probability inequalities and generalized moment problems. The second is to characterize the TTD and then develop reliability measures using the characterized TTD. A number of methods use this strategy, including statistical methods, mathematical programming-based methods, and utility-based methods.
- The modeling framework based on assumed distributions of uncertainty sources also provides two ways for developing reliability measures to evaluate TTR. The first way is to directly develop reliability measures using assumed distributions of the uncertainty sources; game theory is a typical method under this approach. The second is to obtain the characterized TTDs and then to develop reliability measures. Current methods of developing reliability measures after obtaining the characterized TTDs is the same as those in the second way of the first modeling perspective (i.e., the above bullet point).

In early studies, TTR is evaluated as the probability that travel times will remain below acceptable levels, and travel times are obtained using extensive simulations designed to solve traffic assignment problems (e.g., Asakura and Kashiwadani, 1991; Du and Nicholson, 1997; Bell *et al*., 1999). Recently, Sumalee and Watling (2003, 2008) and Ng *et al*. (2011) propose different methods to obtain upper bounds instead of exact probabilities, which improve the computational efficiency of the methods. In addition to these single scalar performance indices or bounds, some studies construct the whole PDF of the total travel time to obtain a complete picture of TTR (Clark and Watling, 2005; Ng and Waller, 2010a). In summary, the reliability measures developed under these two modeling perspectives can be divided into three classes: (1) point-based measures, including probability-based, moment-based, percentile-based, tail-based, and utility-based measures, (2) bound-based measures, and (3) PDF-based measures.

Reliability measures for TTR evaluation are summarized in Table 3, regarding their types, formulae, methods, assessing level, and representative references. In the "assessing level" column, "all" means that the reliability measures can be used for evaluating link TTR, route TTR, and network TTR; "route" or "network" means that the reliability measures can be only used for evaluating route TTR or network TTR. In general, reliability measures are assumed to be valid at all assessing levels unless specified otherwise. In Sections 3.1 and 3.2, we review the methods associated with each of the above two modeling perspectives, provide their general formulas, and discuss the corresponding reliability measures.

Table 3. Travel time reliability measures used for evaluating travel time reliability

| Type | Measure | Formula | Method | Assessing level | References |
|---|---|---|---|---|---|
| Probability-based | Probability | $F\left(\overline{T}\right)=\int_0^{\overline{T}} f\left(x\right)dx$ | Statistic | All | Bell and Iida (1997) and Iida (1999); Lam and Xu (1999); Lei *et al*. (2014); Chen *et al*. (2017) |
| | FR | $1-P\left(T<\left(1+\rho\right)\cdot F^{-1}\left(50\%\right)\right)$ | | All | Lomax *et al*. (2003) |
| | | $R\left(\overline{T}\right)=\exp\left(-\int_0^{\overline{T}}\frac{f\left(x\right)}{R\left(x\right)}dx\right)$ | | All | Al-Deek and Emam (2006); Emam and Al-Deek (2006) |
| | FoC | $P\left(T>\left(1+\rho\right)\cdot F^{-1}\left(50\%\right)\right)$ | | All | FHWA (2009) |
| Moment-based | SD | $\sigma$ | Statistic | All | Chen *et al*. (2003); Fosgerau *et al*. (2008); Carrion and Levinson (2013); Mahmassani *et al* (2013); Yang and Wu (2016) |
| | Inverse SD | $1/\sigma$ | | | Polus (1979) |
| | Variance | $\sigma^2$ | | | Rakha *et al*. (2006, 2010) |
| | CoV | $\sigma/\mu$ | | | Lomax *et al*. (2003); Chen *et al*. (2018) |
| Percentile-based | $\rho$-PTT | $F^{-1}\left(90\%\right)$ or $F^{-1}\left(95\%\right)$ | Statistic | All | FHWA (2009); Arezoumandi (2011) |
| | | $F^{-1}\left(\rho\right)$ | | All | Nie and Wu (2009); Nie (2011) |
| | TTI | $\mu/F^{-1}\left(15\%\right)$ | | All | Pu (2011) |
| | PTI | $F^{-1}\left(95\%\right)/F^{-1}\left(15\%\right)$ | | All | NCHRP (2008) |
| | LTI | $\mu/F^{-1}\left(1-\left(1-\rho\right)/2\right)$ | | Link<br>Route | Kaparias *et al*. (2008) |
| | ELI | $F^{-1}\left((1-\rho)/2\right)/\mu$ | | | |
| | N.A. | $F^{-1}(80\%)-F^{-1}(50\%)$ | | All | Loon *et al*. (2011) |
| | Buffer time | $F^{-1}(95\%)-F^{-1}(50\%)$ | | All | FHWA (2009) |
| | BI | $\dfrac{F^{-1}(95\%)-F^{-1}(50\%)}{F^{-1}(50\%)}$ | | All | Lomax *et al*. (2003); SHRP (2009) |
| | $\lambda^{\text{skew}}$ | $\dfrac{F^{-1}(90\%)-F^{-1}(50\%)}{F^{-1}(50\%)-F^{-1}(10\%)}$ | | All | van Lint *et al*. (2008) |
| | $\lambda^{\text{var}}$ | $\dfrac{F^{-1}(90\%)-F^{-1}(10\%)}{F^{-1}(50\%)}$ | | All | van Lint *et al*. (2008) |
| | N.A. | $\left(F_f^{-1}(90\%)-F_f^{-1}(10\%)\right)(1-\rho_{br}) + \left(F_c^{-1}(90\%)-F_c^{-1}(10\%)\right)\rho_{br}$ | | Route | Tu *et al*. (2012) |
| | (T)TTB | $\min\left\{\overline{T} \mid P\left(T \le \overline{T}\right) \ge \rho\right\}$ | | All | Lo *et al*. (2006); Shao *et al*. (2006a, 2016b); Chen *et al*. (2014a) |
| Tail-based | MI | $\dfrac{\int_{0.8}^{1} F^{-1}(x)\,dx-\mu}{\mu}$ | Statistics | All | Lomax *et al*. (2003); FHWA (2009) |
| | ME(T)TT | $\dfrac{1}{1-\rho}\int_{\rho}^{1} F^{-1}(x)\,dx$ | | All | Chen and Zhou (2010); Chen *et al*. (2011b); Xu *et al*. (2013, 2014, 2017) |
| | TTRR | $\frac{\beta+\gamma}{\alpha}\int_{\frac{\gamma}{\beta+\gamma}}^{1} F^{-1}(x)\,dx$ | | All | Fosgerau (2017) |
| | UA | $\int_{\rho}^{1}\left(F^{-1}(x)-F^{-1}(\rho)\right)dx$ | | All | Zang *et al*. (2021) |
| Bound-based | UBP | N.A. | Lempel-Ziv entropy | All | Li *et al*. (2019) |
| | UB | N.A. | Traffic assignment | Network | Sumalee and Walting (2003, 2008) |
| | | $P(T>\bar{T})\leq UB$ | Probability inequalities | Network | Ng *et al*. (2011) |
| | | N.A. | Generalized moment problem | Network | Ji *et al*. (2019) |
| PDF-based | PDF | $f(T)$ | Moment-based<br>Fourier transform | Network | Clark and Watling (2005)<br>Ng and Waller (2010a) |
| Utility-based | TDC | $\sum_{i}\sum_{\omega}U_{i}^{\omega}\cdot q_{i}^{\omega}$ | Traffic assignment | Network | Yin and Ieda (2001); Yin *et al*. (2004) |
| | LAP | $\eta E[\max(0, T-\bar{T})]$ | | Network | Watling (2006) |
| | Reliability premium | N.A. | Schedule delay | Route | Batley (2007); Beaud *et al*. (2016); Zang *et al*. (2021) |
| | NED | $\mu+\eta E[T-\bar{T}]$ | Schedule delay | All | Lee *et al*. (2019) |

Note: 1. FR: failure rate; FoC: frequency of congestion; SD: standard deviation; CoV: coefficient of variation; PTT: percentile of travel time; TTI: travel time index; PTI: planning time index; LTI: lateness index; ELI: earliness index; BI: buffer index; TTB: travel time budget; MI: misery index; METT: mean-excess travel time; TTRR: travel time reliability ratio; UA: unreliability area; UBP: upper bound of predictivity; UB: upper bound; TDC: total disutility of commuter; LAP: late arrival penalty; NED: normalized expected disutility.

2. N.A. means that (1) it is hard to give a specified name of the reliability measures, or (2) it is hard to state the measure in a simple expression; readers can refer to the original references for details.

3. $F_{f}^{-1}$ is the percentile travel time in free flow conditions; $F_{c}^{-1}$ is the percentile travel time in the congested conditions after breakdown; $\rho_{br}$ is the probability of traffic breakdown.

## 3.1 Evaluating TTR with Travel Time Datasets

### 3.1.1 Evaluating TTR with only moments information of travel times

Note that probability inequalities and generalized moment problem are proposed for evaluating network-wide TTR, so their corresponding travel time variable is total travel time.

*(1) Probability inequalities*

Specifically, the upper bound of the reliability engineering function (also called as the survivor function) of total travel time is used as the reliability measure to evaluate network TTR. Let $R(\bar{T})$ denote the reliability function of total travel time $T$ at a specified threshold $\bar{T}$, i.e., $R(\bar{T}) = P(T \geq \bar{T})$. Then, the upper bound of the reliability function can be written as

$$\max R(\bar{T}) = \max P(T \geq \bar{T}) \tag{13}$$

Using common probability inequalities, including the Markov inequality (Ross, 2002) and Madansky inequality (Madansky, 1959), Ng *et al.* (2011) derives several specific upper bounds of reliability function to assess the network reliability performance, including (1) upper bound based on the first-order moment, (2) upper bound based on both the first and second-order moments, and (3) upper bound based on the first $n$ moments.

*(2) Generalized moment problem*

Similarly, to assess network TTR, Ji *et al.* (2019) proposes a generalized moment problem for obtaining the upper bound of the reliability function $R(\bar{T})$. Assume that $\mathbf{t} = (v_1 t_1, v_2 t_2, \ldots, v_a t_a, a \in A)$, which is the vector of total link travel times; $T = \sum v_a t_a$ is the total travel time, $\Omega_\mathbf{t}$ is the support information about $\mathbf{t}$; $S = \{T \geq \bar{T}, \mathbf{t} \in \Omega_\mathbf{t}\}$; and $\mathbf{1}_S = 1$ if $\mathbf{t}$ is in the set $S$, and $\mathbf{1}_S = 0$ otherwise. Then, the reliability function can be rewritten as

$$R(\bar{T}) = P(T \geq \bar{T}) = \int_S f(\mathbf{t})\,d\mathbf{t} = \int_{\Omega_\mathbf{t}} \mathbf{1}_S f(\mathbf{t})\,d\mathbf{t} \tag{14}$$

The upper bound of the reliability function can be formulated as a semi-definite program with finite constraints based on the given first $n$ moments ($\xi_i$ is the $i$th moment of $\mathbf{t}$)

$$\begin{aligned} &\max \int_{\Omega_\mathbf{t}} \mathbf{1}_S f(\mathbf{t})\,d\mathbf{t} \\ &\text{s.t.} \int_{\Omega_\mathbf{t}} \mathbf{t}^i f(\mathbf{t})\,d\mathbf{t} = \xi_i, \quad \forall i \in \mathbb{N}^n \\ &f(\mathbf{t}) \geq 0 \end{aligned} \tag{15}$$

Ji *et al*. (2019) refers to this semi-definite program as a generalized moment problem and mentions that the upper bounds that they derive are tighter than those in Ng *et al*. (2011).

### 3.1.2 Evaluating TTR with characterized travel time distributions

Note that all of the following evaluation methods yield a single scalar performance index.

*(1) Statistical methods: probability*

According to Asakura and Kashiwadani (1991), Bell and Iida (1997), and Iida (1999), TTR is the probability that a trip can be achieved within a given period. In more recent studies, this given period is interpreted as the upper threshold of travel time accepted by travelers, which consists of two components: expected travel time $\mu$ and additional time $\delta$ (Florida Department of Transportation, 2000). This definition of TTR can be formulated as:

$$R\left(\bar{T}\right)=P\left(T\leq\bar{T}\right)=P\left(T\leq\mu+\delta\right)=\int_{-\infty}^{\mu+\delta}f\left(x\right)dx=F\left(\bar{T}\right) \tag{16}$$

This definition is used by Lam and Xu (1999), Lei *et al*. (2014), and Chen *et al*. (2017), and can be interpreted as the probability of completing a trip successfully. In contrast, some studies (Lomax *et al*., 2003; Al-Deek and Emam, 2006; Emam and Al-Deek, 2006) highlight the importance of the probability of failure rate in evaluating TTR. Furthermore, the frequency of congestion (FHWA, 2009) quantifies the probability that the travel time will exceed a specified expected threshold. Consequently, an alternative definition of TTR is

$$R\left(\bar{T}\right)=P\left(T\geq\bar{T}\right)=1-F\left(\bar{T}\right)=e^{-\int_{0}^{\bar{T}}g(x)dx}=e^{-\int_{0}^{\bar{T}}\frac{f(x)}{R(x)}dx} \tag{17}$$

where $g(x)$ is the failure rate function: $g(x)=f(x)/R(x)$. The definition of TTR given in Eq. (17) is known as the reliability engineering function. Interested readers can refer to Rausand and Hoyland (2003) for more details.

*(2) Statistical methods: moment*

The moments of a function are quantitative measures related to the shape of the function's distribution. According to Carrion and Levinson (2012), TTR is a measure of the spread of TTD, and thus, it is natural to use the moments of travel time to evaluate TTR. Given a continuous TTD, the first moment is the expected travel time $\mu$ and the $i$-th central moment can be obtained by

$$R\left(T\right)=\xi_{i}=\int_{-\infty}^{\infty}\left(T-\mu\right)^{i}f\left(T\right)dT,\ \text{for}\ \ i>1 \tag{18}$$

Among the first $n$ moments of travel time, the two most used measures for assessing the reliability performance of transportation networks are the second central moment (variance) (Rakha *et al*., 2006, 2010) and the associated standard deviation (Chen *et al*., 2003; Fosgerau *et al*., 2008; Carrion and Levinson, 2013; Mahmassani *et al*., 2013). Other moment-based measures include the inverse standard deviation (Polus, 1979) and the coefficient of variation (Lomax *et al*., 2003).

*(3) Statistical methods: percentile*

As extensive empirical studies have demonstrated the asymmetric and right-skewed features of TTDs, the percentile function has received increasing attention. Generally, the formula of the percentile function is

$$R(T) = F^{-1}(\rho) \tag{19}$$

where $\rho$ is the reliability requirement. As shown in Table 3, percentile-based reliability measures include $90^{th}/95^{th}$ percentile (Chen *et al*., 2003; FHWA, 2009, Arezoumandi, 2011) or $\rho$-percentile travel time (Nie and Wu, 2009; Nie, 2011), travel time index, planning time index (NCHRP, 2008), buffer time and buffer index (Lomax *et al*., 2003; SHRP, 2009), lateness index and earliness index (Kaparias *et al*., 2008), and skewness-width $\lambda^{skew}$- $\lambda^{var}$ (van Lint *et al*., 2008). Furthermore, the reliability measure by Tu *et al*. (2012), which combines both the uncertainty and the instability in travel times, can also be viewed as a percentile-based measure.

*(4) Statistical methods: distribution tail*

Empirical studies (Odgaard *et al*., 2005; van Lint *et al*., 2008; Franklin and Karlström, 2009; Sikka and Hanley, 2013) show that the unexpected delays resulting from the distribution tail of travel time can lead to serious consequences. Tail-based reliability measures have been developed to consider the effects of distribution tails in TTR evaluation. Tail-based measures are also referred to as tardy trip indicators, and are used to answer the question "How often will a traveler be unacceptably late?" The key for these measures is modeling the distribution tail, which has the following general formula:

$$R(\rho) = \int_{\rho}^{1} F^{-1}(x)\,dx \tag{20}$$

The classic tail-based measures include the misery index (FHWA, 2009), mean-excess travel time (Chen and Zhou, 2010), travel time reliability ratio (Fosgerau, 2017), and unreliability area (Zang *et al*., 2021). Chen *et al*. (2022a, b) propose a conservative expected travel time measure for disseminating travel time distribution information with an explicit consideration of distribution tail to the public and service platforms.

*(5) Mathematical programming method*

Using mathematical programming to define TTR measures is an important alternative method for TTR evaluation. This method considers travelers' tradeoff between trip efficiency and trip cost and is associated with the behavioral assumption of travelers in considering TTR. In other words, travelers tend to add a safety margin $\delta$ beyond the expected travel time $\mu$ to improve trip TTR (Garver, 1968; Thomson, 1969; Knight, 1974; Hall, 1983; Senna, 1994). The sum of the mean travel time and the specified safety margin is referred to as the effective travel time (Hall, 1983). Mathematically, the optimal $\delta$ in the effective travel time can be determined by the following chance-constrained programming:

$$\begin{aligned} & TTB(\rho) = \min\{\mu + \delta\} \\ \text{s.t.} \quad & \int_{-\infty}^{\mu+\delta} f(t)\,dt \geq \rho \end{aligned} \tag{21}$$

The effective travel time is further conceptualized as the travel time budget in Lo *et al*. (2006), which replaces the safety margin with the travel time margin, i.e., the product of the route's standard deviation ($\sigma$) and travelers' punctuality requirement ($\lambda$): $\delta = \lambda\sigma$. As Wu and Nie (2011) notes, when travel times on all routes of an O-D pair follow the same type of distribution, there is one-to-one correspondence between travel time budget (TTB) and percentile travel time. That is, $TTB(\rho) = \mu + \delta = \mu + \lambda(\rho)\sigma = F^{-1}(\rho)$. As can be seen from Eq. (21), the effective travel time or travel time budget considers the reliability aspect of TTD while ignoring the unreliability aspect. To capture the unreliability aspect of travel time related to excessively late trips, Chen and Zhou (2010) defines the mean-excess travel time (METT) as the conditional expectation of travel times beyond the travel time budget:

$$METT(\rho) = \min\left\{TTB(\rho) + \frac{1}{1-\rho}\int_{\rho}^{1}\left(F^{-1}(t) - TTB(\rho)\right)dt\right\} \tag{22}$$

*(6) Utility-based method*

In utility-based methods, travelers usually have a target travel time or preferred arrival time for their trips. Because of travel time variability, travelers may arrive early or late at the destination, defined as schedule delay early or schedule delay late. To improve trip reliability, travelers attach penalty costs to uncertainty in arrival times. The corresponding utility-based measures mainly include the total disutility of commuters (Yin and Ieda, 2001; Yin *et al*., 2004), late arrival penalty (Watling, 2006), normalized expected disutility (Lee *et al*., 2019), and reliability premium (Batley, 2007; Beaud *et al*., 2016; Zang *et al*., 2021). The expressions of the above utility-based measures can be found in Table 3. The explicit formula for the reliability premium depends on the linear utility function, travelers' risk attitudes, etc. Interested readers can refer to Batley (2007) and Beaud *et al*. (2016) for the expressions of the reliability premium under different assumptions. Note that the theoretical basis of the utility-based method is the valuation models of TTR to be reviewed in Section 4.1.

It should be noted that many reliability measures such as travel time budget and mean-excess travel time can be regarded as route utilities in the route choice model. However, reliability measures are categorized as utility-based measures in this paper only when they are explicitly based on utility theories.

## 3.2 Evaluating TTR with Assumed Distribution of Uncertainty Sources

### 3.2.1 Obtaining the characterized TTDs and then developing reliability measures

We briefly review how to obtain the characterized TTDs based on an assumed distribution of uncertainty from the supply side (Lo and Tung, 2003; Lo *et al*., 2006), the demand side (Shao *et al*., 2006a, 2006b), and both the demand and supply sides (Lam *et al*., 2008).

*(1) Demand-side uncertainty*

Assume that the stochastic demand $\tilde{q}_\omega$ of O-D pair $\omega$ follows an independent theoretical distribution with a mean of $q_\omega$ and a variance of $\left(\sigma_q^\omega\right)^2$. Typically, it is assumed that (1) the route flow denoted by $f_{\omega j}$ follows the same type of probability distribution as the O-D demand; (2) the coefficient of variation (CoV) of the route flow is equal to that of the O-D demand; and (3) the route flows are mutually independent. According to the flow conservation constraint, the mean and standard deviation of the stochastic route flow are

$$\mu_f^{\omega j} = E\left[\tilde{f}_{\omega j}\right] = \rho_{\omega j} E\left[\tilde{q}_\omega\right] = \rho_{\omega j} q_\omega, \ \forall j \in J_\omega, \ \forall \omega \in \Omega_\omega \ \text{ and}$$
$$\sigma_f^{\omega j} = \sqrt{Var\left[\tilde{f}_{\omega j}\right]} = f_{\omega j} CoV_\omega = f_{\omega j} \frac{\sigma_q^\omega}{q_\omega}, \ \forall j \in J_\omega, \ \forall \omega \in \Omega_\omega \tag{23}$$

where $\rho_{\omega j}$ is the choice probability of route $j$ between O-D pair $\omega$. Then, based on the definitional constraint, the mean and standard deviation of the stochastic link flow are

$$\mu_v^a = E\left[\tilde{v}_a\right] = \sum_{\omega \in \Omega_\omega} \sum_{j \in J_\omega} \delta_{aj} E\left[\tilde{f}_{\omega j}\right] = \sum_{\omega \in \Omega_\omega} \sum_{j \in J_\omega} \delta_{aj} \mu_f^{\omega j}, \ \forall a \in A \ \text{ and}$$
$$\sigma_v^a = \sqrt{\sum_{\omega \in \Omega_\omega} \sum_{j \in J_\omega} \delta_{aj} Var\left[\tilde{f}_{\omega j}\right]} = \sqrt{\sum_{\omega \in \Omega_\omega} \sum_{j \in J_\omega} \delta_{aj} \left(f_{\omega j}\right)^2 \left(\sigma_q^\omega / q_\omega\right)^2}, \ \forall a \in A \tag{24}$$

As we have $\tilde{T}_a = t_a\left(\tilde{v}_a, C_a\right)$, the mean and variance of the link travel time can be computed as:

$$\mu_T^a = E\left[\tilde{T}_a\right] = E\left[t_a\left(\tilde{v}_a, C_a\right)\right],\ \forall a \in A \quad \text{and}$$
$$\left(\sigma_T^a\right)^2 = Var\left[\tilde{T}_a\right] = Var\left[t_a\left(\tilde{v}_a, C_a\right)\right],\ \forall a \in A \tag{25}$$

*(2) Supply-side uncertainty*

In this condition, the uncertainty directly propagates from the source to the link cost through a link cost function. Given the link capacity distribution or free-flow TTD, one can derive an explicit expression of the link TTD (i.e., the mean and variance of the link travel time) based on the link travel time function. Assuming that the link capacity is stochastic, we have $\tilde{T}_a = t_a\left(v_a, \tilde{C}_a\right)$, and then the mean and variance of the link travel time can be computed as:

$$\mu_T^a = E\left[\tilde{T}_a\right] = E\left[t_a\left(v_a, \tilde{C}_a\right)\right],\ \forall a \in A \quad \text{and}$$
$$\left(\sigma_T^a\right)^2 = Var\left[\tilde{T}_a\right] = Var\left[t_a\left(v_a, \tilde{C}_a\right)\right],\ \forall a \in A \tag{26}$$

To have more detailed explicit formulas of mean and variance, one can further assume that the link capacity follows a specific distribution (e.g., the uniform distribution). Interested readers may refer to Lo *et al*. (2006) for more details.

*(3) Both demand- and supply-side uncertainties*

Assuming that the link capacity and the demand are both stochastic, we have $\tilde{T}_a = t_a\left(\tilde{v}_a, \tilde{C}_a\right)$, and then the mean and variance of the link travel time can be computed as follows:

$$\mu_T^a = E\left[\tilde{T}_a\right] = E\left[t_a\left(\tilde{v}_a, \tilde{C}_a\right)\right],\ \forall a \in A \quad \text{and}$$
$$\left(\sigma_T^a\right)^2 = Var\left[\tilde{T}_a\right] = Var\left[t_a\left(\tilde{v}_a, \tilde{C}_a\right)\right],\ \forall a \in A \tag{27}$$

With the derived random link travel time and corresponding statistics, the link (or route or network) TTDs can be obtained based on the methods reviewed in Section 2.2.4. With the characterized TTDs, it is straightforward to develop reliability measures, as discussed in Section 3.1.2. Although all of the reliability measures listed in Section 3.1.2 can be computed in theory, the whole PDF (Clark and Watling, 2005; Ng and Waller, 2010a), travel time budget (Lo *et al*., 2006; Shao *et al*., 2006a), and mean-excess travel time (Chen and Zhou, 2010) are some common reliability measures under this modeling perspective.

### 3.2.2 Directly developing reliability measures

Game theory is a representative methodology for directly developing reliability measures based on assumed sources of uncertainty. The source of uncertainty in game theory is the stochastic traffic supplies that are the result of link failures. In particular, compared to the route choice criterion used in traditional user equilibrium or stochastic user equilibrium, the route choice criterion in game theory assumes that the demons select links that will cause the maximum damage to travelers, while travelers accordingly seek the best routes to avoid link failures (e.g., Bell, 2000; Bell and Cassir, 2002; Szeto *et al*., 2006; Szeto, 2011). The resulting trip cost obtained by the game theory model is used as the measure to evaluate the network performance reliability. Please refer to Section 5.2.1 for the general formula of game theory model.

## 3.3 Analysis of TTR Evaluation Measures

This section summarizes the existing analysis of reliability measures from different perspectives, including behavioral assumptions, consistency, and accuracy.

*(1) Behavioral assumptions*

A widely used behavior assumption in the development of reliability measures is that travelers add a safety margin beyond the mean travel time to hedge against travel time variability (Garver, 1968; Thomson, 1969; Knight, 1974; Hall, 1983; Senna, 1994). Specifically, the reliability measures that consider the reliable aspect of travel times use $90^{th}$/$95^{th}$ percentile or $\rho$-percentile travel time, buffer time and buffer index, and travel time budget. The measures that consider both the reliable and unreliable aspects of travel times include misery index, mean-excess travel time, and travel time reliability ratio. Reliability measures that only consider the consequences of the worst trips include failure rate, frequency of congestion, and unreliability area. Besides, Tan *et al*. (2014) examines the Pareto efficiency of four widely used reliability measures for risk-taking behavior in terms of mean and standard deviation of travel times, i.e., the percentile travel time, travel time budget, mean-excess travel time, and quadratic disutility function.

*(2) Consistency analysis*

Most of reliability measures summarized in Table 3 are related to the statistical properties (particularly the shape) of the TTDs. However, this does not mean that these measures behave consistently for the same assessment object. Some studies demonstrate that there are consistencies between some measures. For example, Higatani *et al*. (2009) finds that buffer time and buffer time index have tendencies similar to those of standard deviation and coefficient of variation, respectively. Dowling *et al*. (2009) and Fosgerau (2017) find that standard deviation is a good proxy for several other measures, whereas Pu (2011) demonstrates that coefficient of variation rather than standard deviation is a good proxy for several other measures. In contrast, many studies have verified the inconsistencies between reliability measures. For example, Loon *et al*. (2011) finds that the correlations between six common measures are lower than expected. Similarly, in a comparative study of 11 measures based on users' and operators' viewpoints, Wakabayashi and Matsumoto (2012) shows that TTR measures of the same route give inconsistent outputs. Because of this inconsistency, different studies recommend different TTR evaluation measures. For instance, van Lint *et al*. (2008) and Tu *et al*. (2007) argue that percentile-based measures are more robust for capturing the often wide and (left) skewed TTDs; Chen and Zhou (2010) states that the mean-excess travel time performs better than travel time budget for risk-averse travelers, which is further confirmed by Xu *et al*. (2014) for network-wide reliability assessment and by Zang *et al*. (2021) for valuing the TTR. Wakabayashi and Matsumoto (2012) suggests that TTR evaluation measures should be selected according to the study's purpose and the characteristics of the study route. To understand the fundamental causes of the observed consistencies and inconsistencies, Pu (2011) theoretically examines their mathematical relationships and interdependencies with the assumed lognormal distributed travel time, and Xu *et al*. (2021) theoretically examines the mathematical and behavioral (in)consistency of the schedule delay, travel time budget, and mean-excess travel time models.

*(3) Accuracy analysis*

Some studies examine the accuracy of TTR measures in different applications. Yang and Cooke (2018) develops a bootstrapping technique-based framework to explore the accuracy of TTR measures and finds that moment-based TTR measures are not sufficiently accurate for evaluating freeway TTR. Considering the computational burden, Rakha *et al*. (2006, 2010) propose five methods for calculating trip travel time variance based on link travel time variance and conclude that the accuracy of trip travel time variance can be ensured when calculating the coefficient of variation of trip travel time as the conditional expectation over all of links. To improve the estimation of network travel time reliability, Saedi *et al*. (2020) uses network partitioning to divide the heterogeneous large-scale network into homogeneous regions with well-defined network fundamental diagrams.

## 4. TRAVEL TIME RELIABILITY VALUATION

In this section, we review theoretical models and valuation measures for quantifying the VOR, and then summarize the existing valuation dimensions of TTR.

According to Li *et al*. (2010) and Carrion and Levinson (2012), the VOR reflects the value of the monetary unit that travelers are willing to pay or place to improve the reliability of their travel time. There are four types of theoretical models for quantifying the VOR in transportation networks: mean-variance, schedule delay, mean-lateness, and network utility maximization models. In addition to theoretical models, stated preference and revealed preference surveys can empirically estimate the VOR (see, e.g., Bates *et al*., 2001; Lam and Small, 2001; Bhat and Sardesai, 2006; Liu *et al*., 2004, 2007; Batley and Ibáñez, 2009; Carrion and Levison, 2013). As this paper focuses on the theoretical models, interested readers may refer to Li *et al*. (2010), Carrion and Levison (2012), and Shams *et al*. (2017) for in-depth reviews of the VOR research based on stated preference and revealed preference surveys. Table 4 briefly summarizes the disutility source, theoretical foundation, utility function, model components, and representative references of these four types of models.

The main difference among the four types of models for valuing TTR is their different assumptions regarding the sources of disutility, and the different assumed sources of disutility further affect their model components considered.

- For the mean-variance model, the source of disutility is the whole TTD. Therefore, the associated valuation metrics generally measure the width of the TTD, including both the right-hand and left-hand sides of the TTD, and the mean-variance model consists of two utilities: the utility of mean travel time and the utility of standard deviation.
- For the schedule delay model, the disutility is incurred when the traveler does not arrive at the destination at the preferred arrival time, and thus the corresponding valuation metrics focus on schedule delay[7], including both schedule delay early and schedule delay late. Consequently, the schedule delay model contains utilities of travel time, schedule delay early, and schedule delay late.
- As late arrival is much more serious than early arrival, the mean-lateness model assumes that the disutility only results from late arrival. Thus, the mean-lateness model has only two utility components, i.e., the utility of travel time and the utility of the lateness at arrival.
- In the network utility maximization model, the disutility comes from the utility maximization behavior of the drivers in transportation networks. This model is expressed as a mathematical programming subject to several budget constraints.

[7] Schedule delay is also described as the variations to the preferred arrival time in Bates *et al*. (2001).

Table 4. Approaches used in VOR studies

| Mathematical model | Disutility source | Theoretical foundation | Utility function | Components | References |
|---|---|---|---|---|---|
| Mean-variance | Travel time variability | Expected utility theory | Constant | Travel time and variability measure | Jackson and Jucker (1982)<br>Pells (1987); Black and Towriss (1993) |
| | | | Polynomial or exponential | Travel time and variability measure | Polak (1987) |
| | | | General form | Travel time, variability measure, and cost | Senna (1994); Small *et al*. (1999) |
| Schedule delay | Not arriving at the destination at the preferred arrival time | Utility maximization | Constant | Travel time, SDE, and SDL (lateness penalty) | Small (1982); Noland and Small, (1995); Bates *et al*. (2001); Hollander (2006); Fosgerau and Karlström (2010) |
| | | | Non-constant | Travel time, SDE, and SDL (lateness penalty) | Tseng and Verhoef (2008); Li *et al*. (2010); Fosgerau and Engelson (2011); Jenelius *et al*. (2011) |
| Mean lateness | Mean lateness at departure and/or arrival | Expected utility theory | Constant | Travel time and SDL | ATOC (2005); Batley and Ibáñez (2009, 2012) |
| Network utility maximization | Utility maximization behavior of the drivers in the network | Utility maximization behavior in a network | Cobb–Douglas | Objective function, budget constraint, and driver behavior | Uchida (2014); Kato *et al*. (2020) |

Note: SDE: schedule delay early; SDL: schedule delay late.

In general, the mean-variance model, schedule delay model, and mean-lateness model investigate VOR on the trip level, although some studies extend them to trip chains (e.g., Jenelius *et al.*, 2011; Jenelius, 2012) or bottlenecks (e.g., Siu and Lo, 2009; Coulombel and de Palma, 2014a, 2014b; Zhu *et al.*, 2018). Figure 5 provides an illustration to help readers have a better understanding of these three models, particularly their differences and similarities. In contrast, Uchida's (2014) network utility maximization model estimates the value of time and VOR in transportation networks using the network structure and drivers' route choice behavior.

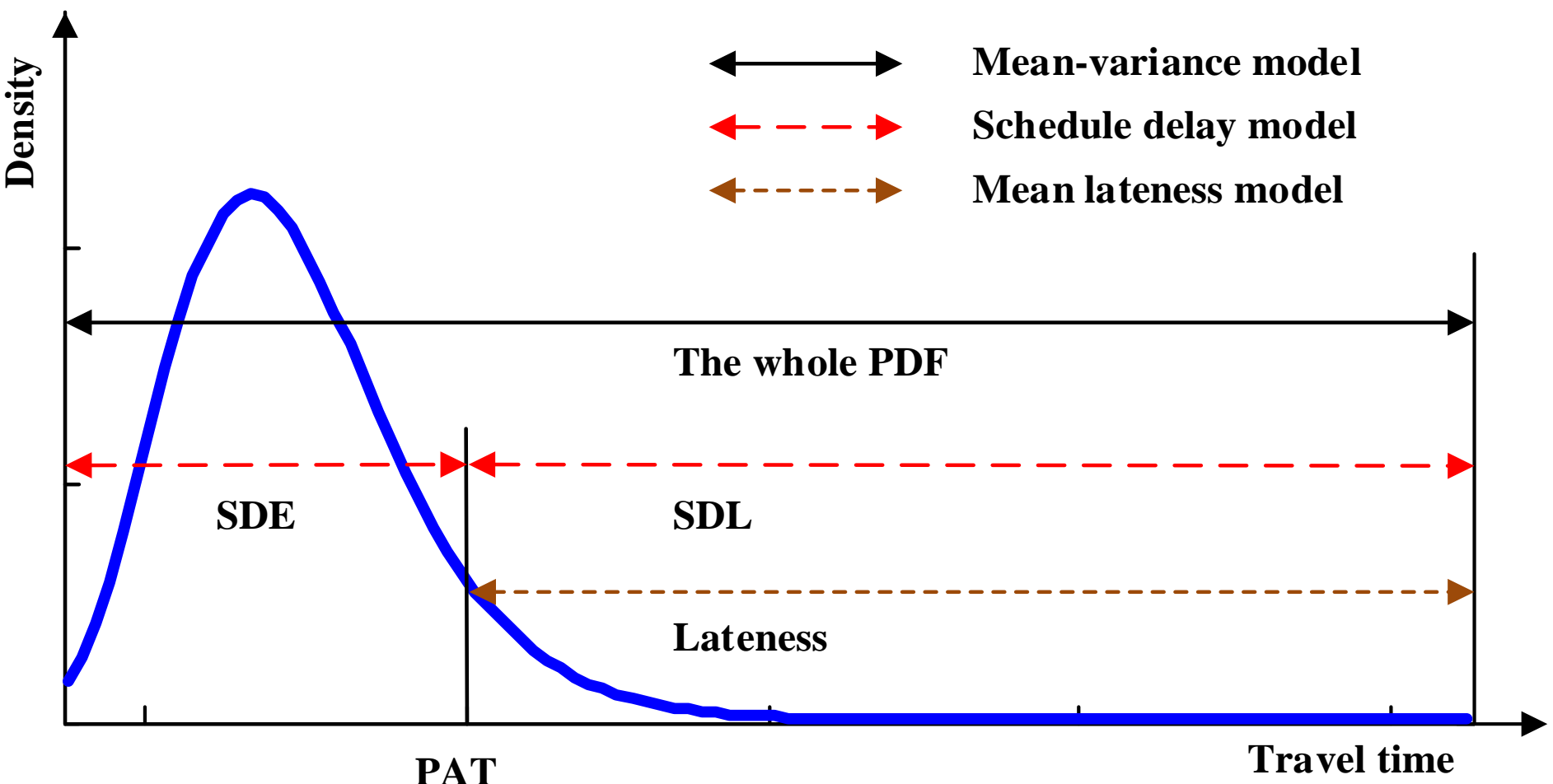

Figure 5. Different sources of utility in the mean-variance, schedule delay, and mean lateness models (PAT: preferred arrival time; SDE: schedule delay early; SDL: schedule delay late)

## 4.1 Theoretical Models for Valuing TTR

This section reviews four theoretical models for valuing TTR. It is important to note that all of the $\eta$ coefficients in the following equations are parameters to be estimated.

### 4.1.1 Mean-variance model

The basic idea of the mean-variance model is that both expected travel time and travel time variability are sources of disutility (Jackson and Jucker, 1982; Pells, 1987; Black and Towriss, 1993), and thus the utility in the mean-variance model is defined as a function of expected travel time and travel time variability. The mean-variance model is sometimes referred to as the Bernoulli approach (e.g., Beaud *et al.*, 2016), as it dates back to the expected utility approach originally proposed by Bernoulli (1738) and perfected by von Neuman and Morgenstern (1947).

In the transportation literature, Jackson and Jucker (1982) first proposed the mean-variance framework to study the traveler's tradeoff between travel time and travel time variability, where travelers minimize the sum of the disutility of travel time and the disutility of travel time variability as follows:

$$U(T) = \eta_1\mu + \eta_2 Var[T] \quad (28)$$

Using the framework in Jackson and Jucker (1982), Polak (1987) considers alternative formulae of utility function to understand the risk behavior of travelers in response to travel time variability, i.e., the polynomial of second degree and exponent form of travel time. By combining the mean-variance model in Jackson and Jucker (1982) and the expected utility approach in Polak (1987), Senna (1994) derives a general algebraic term of degree with respect to the travel time, which makes it possible to measure risk aversion (or proneness). In addition, Senna's model has another attribute: trip cost $c$, i.e.,

$$EU(T) = \eta_1 E[T] + \eta_2 Var[T] + \eta_3 c \quad (29)$$

### 4.1.2 Schedule delay model

Although the mean-variance model assumes a trade-off between the expected travel time and travel time variability for travelers, it ignores the travelers' scheduling cost. To fill in this gap, Small (1982) extends the seminal work of Gaver (1968) and Vickery (1969) on the schedule delay model to include the scheduling cost. In the schedule delay model, the scheduling cost plays a major role in the travelers' departure time choice, and travelers' disutility is due to not arriving at the preferred arrival time, whether early or late. Let *Arr* and *D* denote the arrival time and departure time, respectively. The schedule delay model in Small (1982) is expressed as follows:

$$U(D,T) = \eta_1 T + \eta_2 (SDE) + \eta_3 (SDL) + \eta_4 D_L \quad (30)$$

where *SDE* is the schedule delay early defined as $(PAT - Arr)^+$, and *SDL* is the schedule delay late defined as $(Arr - PAT)^+$. $(\cdot)^+$ denotes a function such that $x^+ = x$ if $x > 0$, and 0 otherwise. $\eta_1$, $\eta_2$, and $\eta_3$ are the traveler's preference parameters, representing the marginal utility per unit of mean travel time, marginal utility per unit of SDE, and marginal utility per unit of SDL, respectively. $D_L$ is a binary term that is equal to 1 if $SDL \geq 0$, and 0 otherwise, and $\eta_4$ is an additional discrete lateness penalty associated with $D_L$.

To account for the effect of different levels of congestion, Noland and Small (1995) extends Small's scheduling model by including the probability distribution of travel time. Therefore, the formula of the expected utility of the scheduling model is

$$EU(D,T) = \eta_1 E[T] + \eta_2 E[SDE] + \eta_3 E[SDL] + \eta_4 D_L \quad (31)$$

For simplicity, the binary lateness penalty can be discarded by assuming that $\eta_4 = 0$, and the corresponding simpler form would be Eq. (32), which is widely used in VOR research (e.g., Fosgerau and Karlström, 2010; Fosgerau and Engelson, 2011; Xiao and Fukuda, 2015; Zang *et al*., 2018b, 2021). Note that $\eta_1$, $\eta_2$, and $\eta_3$ in Eq. (32) are commonly replaced by $\alpha$, $\beta$, and $\gamma$ in the literature.

$$EU(D,T) = \eta_1 E[T] + \eta_2 E[SDE] + \eta_3 E[SDL] \tag{32}$$

Given the above discussion, Bates *et al*. (2001) argues that the simple schedule delay model can be approximated by a mean-variance model when travel time follows assumed distributions. Fosgerau and Karlström (2010) theoretically demonstrates that this equivalence holds as long as the distribution of standardized travel time $X$ is independent of the departure time. Specifically, with the standardized travel time $X = (T - \mu)/\sigma$, the mean-variance model[8] for approximating the optimal expected utility $EU^*$ is the weighted sum of the mean and the product of the standard deviation and mean-lateness factor $H$:

$$EU^* = \eta_1 \mu + (\eta_2 + \eta_3) H\left(\Phi, \frac{\eta_2}{\eta_2 + \eta_3}\right)\sigma \tag{33}$$

where $H\left(\Phi, \frac{\eta_2}{\eta_2 + \eta_3}\right) = \int_{\frac{\eta_2}{\eta_2+\eta_3}}^{1} \Phi^{-1}(\rho)\, d\rho$, and $\Phi^{-1}(\rho)$ is the inverse CDF of standardized travel time $X$.

In Small's scheduling model, the utility function takes a piece-wise constant form, and thus the corresponding schedule delay model is also referred to as the step model. Based on Vickrey (1973), recent studies (Tseng and Verhoef, 2008; Fosgerau and Engelson, 2011) generalize Small's scheduling model with time-dependent preference parameters, and thus, the utility function takes a linear form, which is referred to as the slope model. Figure 6 gives the diagrams of the utility functions in the slope model and the step model, where $h(t)$ is the utility of staying at home and $w(t)$ is the utility of staying at work at time $t$. The analytical expression of the slope model with respect to the linear utility function is

$$U = \int_0^D h(t)dt + \int_{Arr}^0 w(t)dt = \int_0^D (\eta_1 + \eta_2 t)dt + \int_{Arr}^0 (\eta_3 + \eta_4 t)dt \tag{34}$$

Many studies (e.g., Fosgerau *et al*., 2008; Fosgerau and Karlström, 2010; Fosgerau and Engelson, 2011; Fosgerau, 2017) demonstrate that the step model is preferable for theoretically quantifying VOR due to its simple structure, and it is also useful for empirical studies when the preferred arrival time is observable; however, the slope model provides a better explanation of observed scheduling behavior because of its time-varying utility rate, and its related valuation measure (namely travel time variance) is additive over parts of a trip. In addition to the linear utility function shown in Figure 6, non-linear utility functions have been proposed for different purposes, e.g., accounting for risk behavior in Li *et al*. (2012) and Hensher *et al*. (2013) and evaluating VOR for trip chains in Jenelius *et al*. (2011) and Jenelius (2012).

---

[8] Börjesson et al. (2012) refers to the approximated mean-variance model as the reduced form of the scheduling model and further points out that such an approximation may not capture all of the variability costs. This conclusion has been later verified by Xiao and Fukuda (2015) and Abegaz *et al*. (2017).

Note that the schedule delay model characterizes congestion as an exogenous phenomenon and thus fails to capture the impact of travelers' departure time choice on travel time variability/congestion. The trip scheduling equilibrium model based on the bottleneck model addresses this shortcoming (Siu and Lo, 2009; Coulombel and de Palma, 2014a, 2014b; Zhu *et al*., 2018). Although the trip scheduling equilibrium model uses the same framework of the schedule delay model, it takes into account the equilibrium mechanisms/interactions between individuals' departure time choice and congestion. Under the trip scheduling equilibrium model, the expected cost $EC$ for individual $i$ is

$$EC = \frac{Q(i)}{C_b} + E(RD) + \eta_1 E\left[SDE(i)\right] + \eta_2 E\left[SDL(i)\right] \quad (35)$$

where $Q(i)$ is the queue length when individual $i$ entering the bottleneck with the capacity $C_b$ and $RD$ is the random delay.

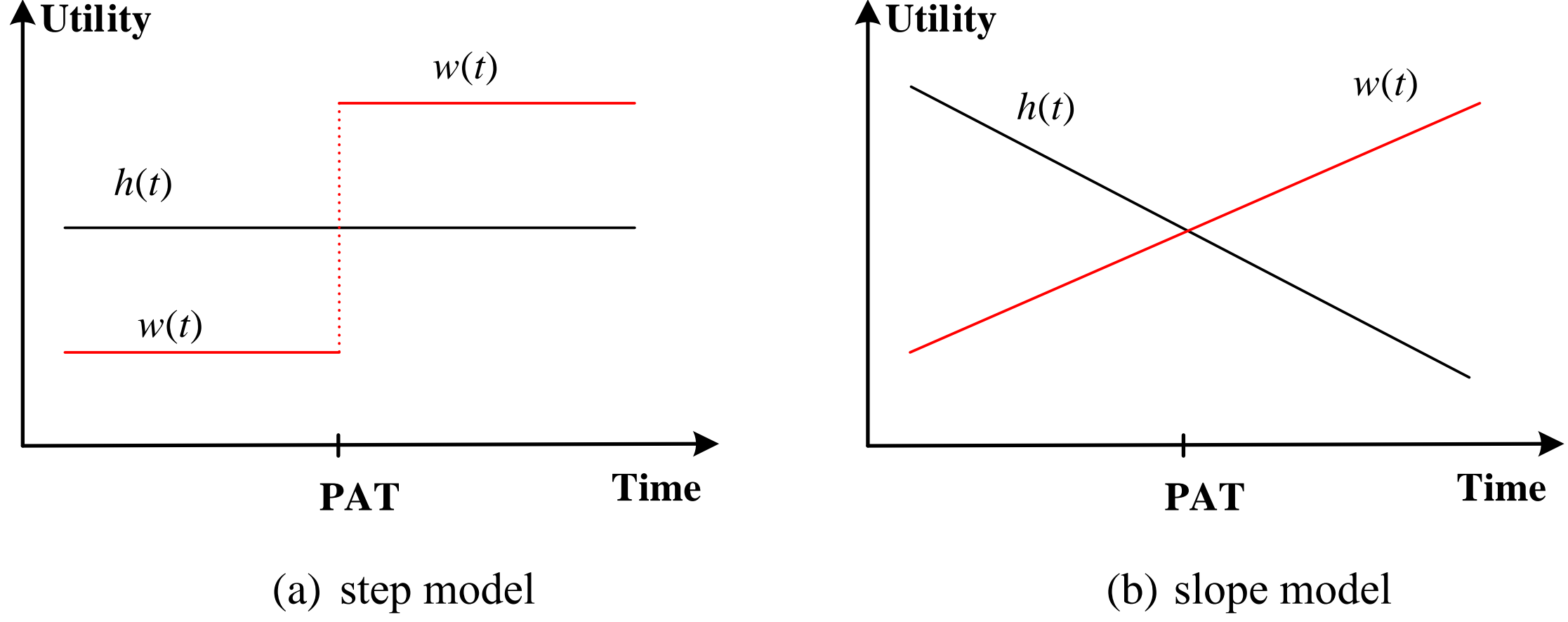

(a) step model (b) slope model

Figure 6. The utility functions of staying at home $h(t)$ and at work $w(t)$ at time $t$ in the (a) step model and (b) slope model (PAT: preferred arrival time)

### 4.1.3 Mean-lateness model

As another approach to measuring VOR, the mean-lateness model considers only the mean lateness at departure and/or arrival and does not consider mean earliness (i.e., negative lateness).

The original model proposed by the Association of Train Operating Companies (2005) is shown in Eq. (36). It describes the reliability of passenger rail transport in the UK and contains two elements under the expected utility paradigm: scheduled journey time and lateness.

$$EU = \eta_1 SchedT + \eta_2 L^+ \tag{36}$$

The scheduled journey time *SchedT* is defined as the travel time between the scheduled departure time and scheduled arrival time; lateness, and the lateness $L^+$, is the difference between the actual arrival time and scheduled arrival time (i.e., the lateness at destination). Batley and Ibáñez (2009, 2012) extend this model by replacing the original lateness $L^+$ with both the lateness at boarding $LB^+$ (i.e., the difference between the actual departure time and scheduled departure time) and the lateness at destination $LD^+$. For more details about the applications and developments of lateness, please refer to Batley *et al*. (2011) and Wardman and Batley (2014).

### 4.1.4 Network utility maximization model

To estimate the value of travel time and VOR in transportation networks, Uchida (2014) formulates a utility maximization problem under three budget constraints related to the travel time $T$, travel cost $c$, and travel time variance $\sigma^2$:

$$\begin{aligned}
&\max \ U_{cd} = \sum_{\omega\in\Omega_\omega} \int_0^{q_\omega} \frac{\lambda_\omega}{x+1} dx = \sum_{\omega\in\Omega_\omega} \lambda_\omega \ln\left(q_\omega + 1\right) \\
&\text{s.t.} \quad \sum_{a\in A} \int_0^{v_a} t_a\left(x\right) dx \le T_{\max} \\
&\qquad \sum_{a\in A} \int_0^{v_a} c_a\left(x\right) dx \le c_{\max} \\
&\qquad \sum_{a\in A} \int_0^{v_a} \sigma_a^2\left(x\right) dx \le \sigma_{\max}^2 \\
&\qquad q_\omega + e_\omega = Q_\omega, \ \forall \omega \in \Omega_\omega
\end{aligned} \tag{37}$$

where the utility $U_{cd}$ to be maximized follows the style of the Cobb–Douglas utility function (Cobb and Douglas, 1928), $\lambda_\omega$ (>0) is a parameter, and $\sum_\omega \lambda_\omega = 1$. $e_\omega$ and $Q_\omega$ are the excess demand and assumed maximal traffic demand, respectively. Let $\varphi_1$, $\varphi_2$, and $\varphi_3$ denote the optimal Lagrangian multipliers associated with the three budget constraints on $T$, $c$, and $\sigma^2$. This utility maximization problem can then be reformulated as a mathematical program that has the same structure as the UE traffic assignment problem with elastic demand:

$$\min \ Z = \sum_{a\in A} \int_0^{v_a} \chi_a\left(x\right) dx - \sum_{\omega\in\Omega_\omega} \int_0^{q_\omega} \frac{\lambda_\omega}{x+1} dx \tag{38}$$

s.t. $f_{\omega j} = \rho_{\omega j} q_{\omega} \quad \forall j \in J_{\omega}, \ \forall \omega \in \Omega_{\omega}$

$$v_a = \sum_{\omega \in \Omega_{\omega}} \sum_{j \in J_{\omega}} \delta_{aj} f_{\omega j}, \ \forall a \in A$$

$$q_{\omega} + e_{\omega} = Q_{\omega}, \ \forall \omega \in \Omega_{\omega}$$

where $\chi_a(v_a)$ is a monotonic increasing function of $v_a$ about link $a$, and its expression is

$$\chi_a(v_a) = \varphi_1 t_a(v_a) + \varphi_2 c_a(v_a) + \varphi_3 \sigma_a^2(v_a) \tag{39}$$

Then, the value of travel time (VOT) and the VOR can be given by

$$VOT = \frac{\varphi_1}{\varphi_2} \text{ and } VOR = \frac{\varphi_3}{\varphi_2} \tag{40}$$

The above network model is derived under the assumption that link travel times are monotonic and separable; interested readers may refer to Uchida (2014) for the network model in which link travel times are non-monotonic and non-separable. In addition, Kato *et al*. (2020) extends the model to consider heterogeneous drivers.

## 4.2 Valuation Measures and Reliability Ratio

Generally speaking, all of valuation measures used to quantify VOR belong to the category of reliability measures summarized in Table 3 of TTR evaluation, because valuation measures need to be able to quantitatively evaluate TTR. However, only a small subset of the TTR evaluation measures are appropriate for valuing TTR, and the existing valuation measures used for empirical studies and theoretical studies of VOR can be divided into two types. For the expressions of the valuation measures described below, please refer to Table 3.

The first type of valuation measures assesses the spread of TTD, but different measures capture such spread from different aspects.

- Standard deviation (e.g., Jackson and Jucker, 1982; Senna, 1994; Hollander, 2006; Batley *et al*., 2008; Fosgerau and Karlström, 2010; Carrion and Levinson, 2013; Coulombel and de Palma, 2014a, 2014b; Zhu *et al*., 2018)
- Variance (e.g., Fosgerau and Engelson, 2011; Uchida, 2014)
- Coefficient of variation (Abdel-Aty *et al*., 1995; Small *et al*., 1995)
- Inter-quantile difference (e.g., Lam and Small, 2001; Brownstone and Small, 2005; Small *et al*., 2005)
- Width of TTD (Hensher, 2001)
- Unreliability area (Zang *et al*., 2021)

Standard deviation, variance, and coefficient of variation can be viewed as moment-based measures, which mainly measure the width of the spread of travel times with respect to the mean and are suitable for characterizing a symmetrical distribution. Compared to standard deviation, variance is additive and thus is convenient for network modeling purposes[9]. Inter-quantile difference and width of TTD can be considered as percentile-based measures, which quantify the reliability of the TTD under a specified confidence level. The unreliability area (Zang *et al*., 2021) focuses on unexpected delays in the tail (i.e., unreliable aspect) of TTD.

The second type of valuation measures uses delays, defined as the difference between the arrival time and a specified value, e.g., preferred arrival time, the most frequently encountered travel time (usual travel time), or timetable. These measures belong to a subset of the utility-based measures summarized in Table 3, including the following.

- Delay early or late (Hollander, 2006, Tilahun and Levinson, 2010)
- Lateness factor (e.g., Franklin and Karlstrom, 2009)
- Reliability premium (e.g., Batley, 2007; Beaud *et al*., 2016; Zang *et al*., 2021)

Based on these mathematical models and valuation measures, the outputs of many VOR studies are the estimated VOT and VOR. However, it is meaningless to directly compare the estimated values of VOT and VOR because studies are done in different areas and use different model structures and different valuation measures. To make a fair comparison among different VOR studies, it is necessary to use the so-called reliability ratio, defined as the ratio of the VOR to the VOT (Jackson and Jucker, 1982; Black and Towriss, 1993). Its expression is

$$Reliability\ ratio = \frac{\partial U / \partial \sigma_T}{\partial U / \partial \mu_T} = \frac{VOR}{VOT} \tag{41}$$

where $\mu_T$ and $\sigma_T$ are the valuation measures for the VOT and VOR, respectively. The reliability ratio can indirectly reveal the importance of reliability relative to the expected time, and hence many reviews use the estimated reliability ratios in VOR studies to investigate how large the VOR is, such as Chang (2010), Li *et al*. (2010), Carrion and Levinson (2012), de Jong *et al*. (2014), and Batley *et al*. (2019).

[9] Engelson and Fosgerau (2011) shows that the variance is a special case of a more general travel time cost measure, i.e., cumulant generating function measure.

### 4.3 Dimensions of Valuing Travel Time Reliability

Based on the well-established schedule delay model, Bates *et al*. (2001) develops a general model of the VOR for personal travel, which brought on huge empirical and theoretical studies of the VOR. Building on this work, Fosgerau and Karlström (2010) derives the simple expression of the optimal expected utility shown in Eq. (33), providing a foundation for subsequent VOR research. In addition to the standard dimension of VOR (e.g., Bates *et al*., 2001; Fosgerau and Karlström, 2010; Fosgerau and Engelson, 2011), there are a variety of dimensions of VOR. Below, we briefly introduce these valuation dimensions.

*(1) The value of service (headway) and service reliability*

Scheduled services, e.g., public transportation, train, or air transportation, usually have long service intervals (i.e., headway), which constrain users' scheduled departure times. To investigate the impact of headway on users' choice, Fosgerau (2009) defines the value of headway as the marginal increase of users' cost with the increased headway. Benezech and Coulombel (2013) extends Fosgerau's work for variable headways and gives two alternative definitions: the value of service and the value of service reliability. These two values quantify the marginal effect of a change in the mean and standard deviation of headway on users' expected travel cost, respectively. Mathematically, a main difference between deriving the value of service or service reliability and deriving the VOR is that the travel time $T$ in the schedule delay model is replaced by the sum of waiting time $T_w$ and in-vehicle time $T_v$.

*(2) The value of information*

Empirical evidence shows that disseminating information of TTR to travelers helps them to make better travel choices (Ettema and Timmermans, 2006; Zhang and Levinson, 2008; Zhu and Timmermans, 2010). To quantify the benefits to users, it is straightforward to define the value of information as the difference between trip utility with and without information (e.g., Engelson and Fosgerau, 2020), i.e., $\left|U\left(D^{*},\ T\right)-U\left(D_{info}^{*},\ T\right)\right|$. $D^*$ and $D_{info}^{*}$ are the optimal departure time for a trip without and with information, respectively. Except for the value of information derived from a better departure time choice, other studies show that the value of information can be also derived from a better route choice (de Palma and Picard, 2006; Gao *et al*., 2010; de Palma *et al*., 2012) and a better departure time and route choice (Soriguera, 2014). In addition, to accurately quantify the value of information, it is necessary to include some critical impact factors in the calculation, such as the cost of obtaining information (e.g., Chorus *et al*., 2006; Fosgerau and Jiang, 2019; Jiang *et al*., 2020), the regimes of information release (de Palma *et al*., 2012; Lindsey *et al*, 2014), and the accuracy of the provided information (Jenelius *et al*., 2011; Engelson and Fosgerau, 2020).

*(3) The cost of misperceived travel time variability*

Xiao and Fukuda (2015) derives the misperceived cost of travel time variability by considering how travelers misperceive TTD. To capture the misperceived cost, a general scheduling delay model is formulated based on the rank-dependent utility theory. Under this modeling framework, the misperceived TTD makes the departure time of travelers deviate from the optimal departure time $D^*$, resulting in a suboptimal departure time $D_{misp}^*$. This deviation allows to quantify the cost of misperceived travel time variability, which is defined as the difference between the expected trip utility under the optimal departure time and the expected trip utility under the suboptimal departure time, i.e., $\left|EU\left(D^*,\ T\right)-EU\left(D_{misp}^*,\ T\right)\right|$. Xiao and Fukuda (2015) presents two analytical expressions of the cost of misperceived travel time variability based on the step utility function and slope utility function, respectively.

*(4) The value of unreliability*

In most VOR studies using the standard scheduling model, the two most widely used measures are the standard deviation and travel time variance (e.g., Hollander, 2006; Fosgerau and Karlström, 2010; Fosgerau and Engelson, 2011; Uchida, 2014; Coulombel and de Palma, 2014a, 2014b; Fosgerau, 2017; Zhu *et al*., 2018). However, the existence of highly-skewed TTDs with long/fat tails has been verified by considerable empirical evidence and thus unexpected delays could have more serious consequences than expected or modest delays (van Lint and van Zuylen, 2005; Fosgerau and Fukuda, 2012; Susilawati *et al*., 2013; Delhome *et al*., 2015; Kim and Mahmassani, 2015; Taylor, 2017; Zang *et al*., 2018b; Li, 2019). To address this phenomenon, Zang *et al*. (2021) distinguishes the value of unreliability from the VOR to capture the unreliability cost caused by the long fat tails of real TTDs. Travelers who consider travel time unreliability will depart early to avoid the serious consequences of unexpected delays, leading to a new departure time $D_{adjust}$. The cost of this adjustment, $|EU(D_{adjust}, T)| - |EU(D^*, T)|$, is the unreliability cost that the traveler must pay to allow for travel time unreliability in the departure time choice. Therefore, the value of unreliability can be defined by the unreliability cost with respect to the unreliability measure.

*(5) The marginal social cost of travel time reliability*

Although the above definitions of VOR reflect how different foci affect the valuation of TTR, their calculations are similar. That is, under all of these extended definitions, VOR can be derived through the difference between utilities under different situations, namely the difference between utilities with the original optimal departure time and the new or adjusted departure time. As discussed above, many factors may affect the departure time, such as service headway, provided information, travelers' misperception of travel time variability, etc. These adjustments to departure time affect and even improve TTR. However, the feedback effect of this adjustment on the congestion profile has been neglected in the above definitions of VOR. To this end, some studies take the congestion as an endogenous factor rather than an exogenous factor to model the interactions between individuals' departure time choice and congestion (e.g., Siu and Lo, 2009; Coulombel and de Palma, 2014a, 2014b; Xiao *et al*., 2017; Zhu *et al*., 2018). The general formula of this equilibrium trip scheduling is given in Eq. (35). The estimated VOR obtained by this equilibrium trip scheduling model is thus interpreted as the marginal social cost of TTR (Coulombel and de Palma, 2014a).

## 5. Travel Time Reliability-Based Traffic Assignment

This section summarizes the methodologies for modeling traffic assignment with consideration of TTR, which will be referred to as TTR-based assignment for short. The general framework of TTR-based assignment is presented in Figure 7.

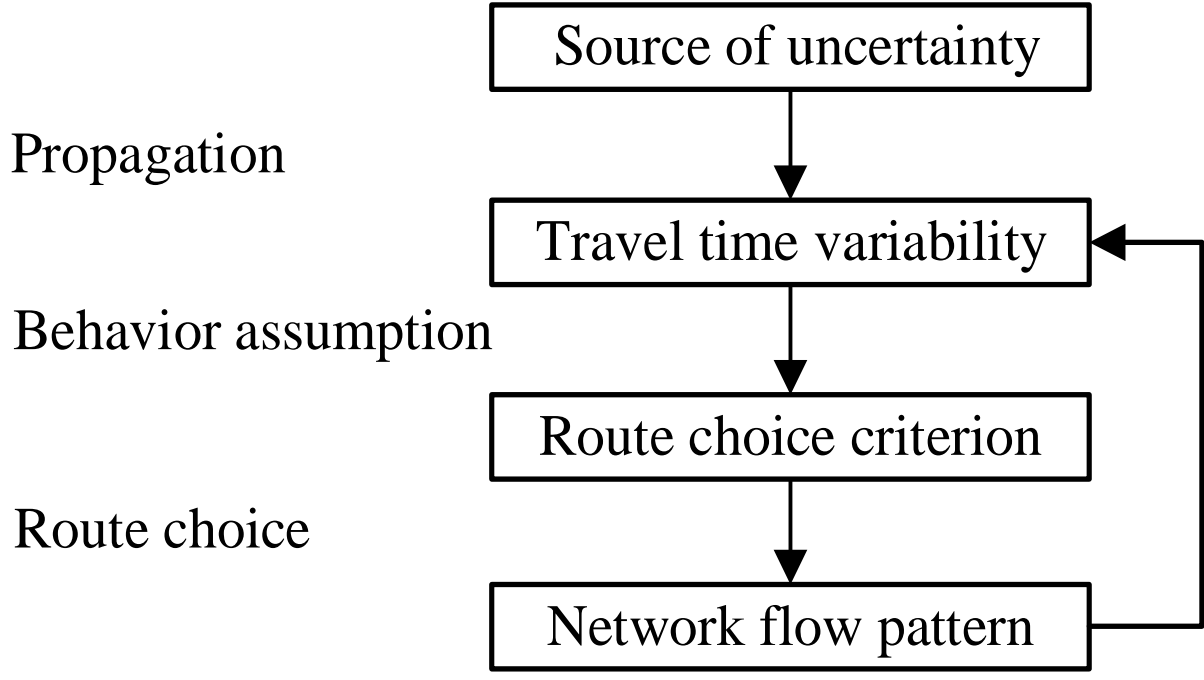

Figure 7. Framework for TTR-based traffic assignment models

As TTR affects individual travelers' route choice behaviors and the collective network flow pattern, TTR-based assignment models have received abundant attention. In these studies, the uncertainty source of travel time variability is typically first specified as coming from the supply side, demand side, or both. Then, based on these assumptions, the uncertainty propagates from its source to the travel time, resulting in travel time variability. A specified route choice criterion for capturing traveler's risk attitude toward such variability is selected for loading the travel demand onto the network. As the uncertainty propagation from the source to the travel time is discussed in Section 2.2.4, the following discussion only introduces the route choice criterion with consideration of TTR, the corresponding traffic assignments models, and their solution algorithms.

Before introducing how to capture travelers' attitudes to TTR in the route choice criterion, it is worthwhile to point out that in the TTR-based traffic assignment, the route choice criterion is used to determine the optimal reliable path for loading the travel demand onto the network. The procedure of determining the optimal reliable path corresponds to the reliable path finding problem in the literature, which is an active research problem conducted by many researchers from different aspects. For example, Miller-Hooks and her colleagues propose several efficient procedures for finding the reliable paths with the least expected time as reliability measure in stochastic and time-varying networks (Miller-Hooks and Mahmassani, 1998, 2000, 2003; Miller-Hooks, 2001; Opasanon and Miller-Hooks, 2006). Shahabi *et al*. (2013, 2015) discuss the robustness of the reliable path finding problem and design solution algorithms for this problem. The Lagrangian relaxation-based algorithms (Xing and Zhou, 2011; Khani and Boyles, 2015; Yang and Zhou, 2017) and simulation-based method (Zockaie *et al*., 2014) are developed for handling link travel time correlation as such correlation cannot be neglected in modeling path travel time reliability. As for the computational efficiency, several algorithms are proposed, including the iterative learning approach in Fakhrmoosavi *et al*. (2019), the Lagrangian substitution algorithm in Zhang and Khani (2019), and the online shortest path algorithm in Khani (2019).

### 5.1 Route Choice Criterion with Consideration of Travel Time Reliability

There are two dimensions considered in capturing travelers' attitudes to TTR: objective quantification and subjective perception. Similar to Chen *et al*. (2002a), Table 5 classifies TTR-based traffic assignment models according to these two dimensions. Objective quantification means transforming the abstract TTR (i.e., characterized TTD) into an intuitive risk measure/utility, as it is difficult to compare different routes using only their TTDs, whereas risk measures in the form of scalars make such comparisons convenient. Obviously, the risk or utility measures used in the TTR-based route choice criterion [10] are the measures for quantitatively assessing TTR and are thus a subset of reliability measures of TTR evaluation summarized in Table 3. Subjective perception reflects how travelers perceive the TTR and can be divided into deterministic perception error and stochastic perception error. Below, we review how these two dimensions are incorporated into the TTR-based route choice criterion.

---

[10] In the literature, TTR-based route choice criterion may be also called as route choice criterion under uncertainty.

Table 5. Classification of TTR-based traffic assignment models

| | | Subjective perception | |
|---|---|---|---|
| | | No | Yes |
| Objective quantification | No | User equilibrium | Stochastic user equilibrium |
| | Yes | TTR-based user equilibrium | TTR-based stochastic user equilibrium |

#### 5.1.1 Dimension 1: objective quantification of travel time reliability

To account for TTR in traffic assignment, various risk or utility measures are used to capture the role of travelers' attitudes toward TTR in route choice criterion. Table 6 summarizes the representative types and the corresponding measure. The first type is based on utility and mainly consists of expected utility-based and non-expected utility-based measures. Expected utility-based measures assume that decision-making under uncertainty aims to maximize the expected utility, e.g., mean-variance (Jackson and Jucker, 1982), late arrival penalty (Watling, 2006), quadratic utility function (Mirchandani and Soroush, 1987), and exponential utility function (Tatineni *et al*., 1997; Chen *et al*., 2002a). In the above utility measures, travelers' different risk attitudes toward TTR can be determined with different parameters. In contrast, the non-expected utility-based measures are based on the idea that decision-making under uncertainty may not be made according to the expected utility maximization, as demonstrated in the cumulative prospect theory (e.g., Avineri, 2006; Connors and Sumalee, 2009; Xu *et al*., 2011), regret theory (Chorus, 2012; Li and Huang, 2017), and fuzzy set theory (Miralinaghi *et al*., 2016). Cumulative prospect theory assumes that travelers compare a summary statistic of the perceived utility (instead of the expected utility) of each route before choosing one. Regret theory uses the modified utility function to calculate route utility and assumes that a traveler can anticipate the possibility that the chosen route may be slower or faster than a non-chosen route, causing regret or satisfaction, respectively. Finally, fuzzy set theory uses the membership value function to calculate route utility, where the mode of the random travel time of each link has the highest membership value and, accordingly, the highest route utility.

Another type of route choice criterion is based on safety margin-based risk measures, which assumes that beyond the mean travel time, travelers add a safety margin to improve TTR (Garver, 1968; Thomson, 1969; Knight, 1974; Hall, 1983; Senna, 1994). Such safety margin-based risk measures include travel time budget (Lo *et al*., 2006; Shao *et al*., 2006a, 2006b; Siu and Lo, 2008), percentile travel time (Nie, 2011; Ordóñez and Stier-Moses, 2010; Wu and Nie, 2013), and mean-excess travel time (Chen and Zhou, 2010; Chen *et al*., 2011b; Xu *et al*., 2017). The former two consider the reliability aspect of travel time variability, and the third one explicitly considers both the reliability and unreliability aspects of travel time variability in the route choice decision process.

Table 6. Route choice criterion under travel time variability

| Types | Risk measures | References | Basic idea |
|---|---|---|---|
| Expected utility-based | Mean-variance | Jackson and Jucker (1982); de Palma and Picard (2005); Di *et al*. (2008); Sumalee *et al*. (2011); Nikolova and Stier-Moses (2014); Prakash *et al*. (2018) | A linear combination of mean and standard deviation of travel time. |
| | Late arrival penalty | Watling (2006) | The route disutility incorporates both the standard generalized travel time and the travel time acceptability in the form of a lateness penalty. |
| | Quadratic utility function | Mirchandani and Soroush (1987); Yin *et al*. (2004); Zhang *et al*. (2011) | A quadratic disutility function that can depict travelers' avoidance attitude toward risk with different parameters. |
| | Exponential utility function | Tatineni *et al*. (1997); Chen *et al*. (2002a) | An exponential disutility function that can reflect different travelers' risk attitude with different parameters. |
| Non-expected utility-based | Perceived value or prospect | Avineri (2006); Connors and Sumalee (2009); Sumalee *et al*. (2009); Xu *et al*. (2011); Yang and Jiang (2014) | A route's prospect value is calculated by the reference point, value function, and weight function. |
| | Expected modified utility (regret) | Chorus (2012); Li and Huang (2017) | A route's utility (regret) is determined by its performance difference with the competing route. |
| | Fuzzy membership function | Miralinaghi *et al*. (2016) | Travelers' perceptions of available routes are described as fuzzy sets and their final |
| | | | decisions are derived from fuzzy set comparison based on some metrics. |
| Safety margin-based | Travel time budget | Lo *et al*. (2006); Shao *et al*. (2006a, 2006b); Siu and Lo (2006, 2008); Lam *et al*. (2008); Chen *et al*. (2011c) | The sum of mean travel time and a safety margin. |
| | Mean-excess travel time | Chen and Zhou (2010); Xu *et al*. (2014, 2017, 2018) | The conditional expectation of travel times beyond the travel time budget. |
| | Percentile travel time | Nie (2011); Wu and Nie (2013) | The percentile of travel time. |
| Others | Robust shortest route | Ordonez and Stier-Moses (2010) | The route that has the best worst-case travel time without being overly conservative by considering a reasonable estimate for the maximum deviation of the route travel time. |
| | Game theory-based measure | Bell (2000); Bell and Cassir (2002); Szeto *et al*. (2006) | Demons select links to cause the maximum damage to travelers, while travelers seek the best routes to avoid link failures. |

There are also some risk measures in the route choice criteria that do not fall into the above two types. For example, in the game theory model in Bell and Cassir (2002), travelers make route choice decisions based on the expected cost triggered by the uncertain disruption caused by the demon. The expected cost therein is different from that in the Wardrop equilibrium, as the link cost is in a binary state (one value in the normal state and another in the disrupted state) rather than following a distribution. Ordonez and Stier-Moses (2010) assumes that travelers will select the robust shortest route with the minimum worst-case travel time. Wu and Nie (2011) uses stochastic dominance to model heterogeneous risk-taking route choice behavior, and Wang *et al*. (2014) assumes that travelers seek routes based on an explicit bi-objective criterion, i.e., minimizing the expected travel time and the standard deviation of travel time.

### 5.1.2 Dimension 2: subjective perception error of travel time reliability

After determining the risk measure, to assign travelers to the network, it is necessary to consider travelers' subjective perception error in the route choice criterion. There are three types of assumptions: no perception error, deterministic perception error, and stochastic perception error.

These three types of assumptions lead to different route cost structures as shown in Figure 8.

- In the first type, it is assumed that travelers have perfect knowledge about the actual route TTD and can choose the optimal route based on some risk measure imposed on the actual route TTD. The route cost is the actual travel cost, denoted by $c$, and thus is a deterministic value that does not have perception error. This leads to the extension of the user equilibrium traffic assignment model (Wardrop, 1952). Representative studies include the probabilistic user equilibrium in Lo and Tung (2003), the travel time budget-based user equilibrium in Lo *et al*. (2006), the late arrival penalized user equilibrium in Walting (2006), the mean-excess travel time-based user equilibrium in Chen and Zhou (2010), and the multi-class percentile user equilibrium in Nie (2011).
- In the second type, it is assumed that travelers' route choice decisions are based on the perceived route TTD, where the perception error is independent of the stochastic travel time. Consequently, the route cost is the perceived cost $c_p$, which is the sum of the actual travel cost $c$ and a deterministic perception error $\varepsilon$ that is independent of $c$, i.e., $c_p = c + \varepsilon$. This assumption is more realistic, as travelers may have imperfect knowledge of the TTD, particularly in congested networks. Accordingly, Siu and Lo (2006) extends the probabilistic user equilibrium model to the stochastic travel time budget equilibrium model. Similarly, Shao *et al*. (2006b) proposes the travel time budget-based stochastic user equilibrium model that is also used by Lam *et al*. (2008), Shao *et al*. (2008), and Siu and Lo (2008). Note that the above models use the common Gumbel variate adopted in Logit-based stochastic user equilibrium model (Dial, 1971; Fisk, 1980) as the perception error term to construct the perceived travel time budget. Recently, Xu *et al*. (2021a) develops a Weibit-based mean-excess travel time stochastic user equilibrium model in the context of day-to-day dynamics, where the perception error follows the Weibull distribution.

- In the third type, travelers' perception error is dependent on the actual travel time distribution, and is thus called "stochastic perception error". The perceived travel cost $c_p(T_p)$ is a function of the perceived travel time $T_p$, which is the sum of the actual TTD and a stochastic perception error conditional on the actual TTD (i.e., $T_p = T + \varepsilon|T$). In proposing this idea, Mirchandani and Soroush (1987) argues that travelers' route choice decisions are based on the perceived TTD rather than on the actual TTD. Chen *et al*. (2011b) proposes a stochastic mean-excess traffic equilibrium model to account for the stochastic perception error. Xu *et al*. (2013) and Wang and Sun (2016) also assume a stochastic perception error.

In addition to the above three types, Walting (2002) proposes a generalized stochastic user equilibrium model to account for day-to-day uncertainty in traffic flows caused by stochastic variation in both the demand and the route choice proportions, conditional on the demands. Different from the conventional stochastic user equilibrium models, in which the perceived cost is made up of the cost at mean flows and perception error, Walting (2002) adds a third part, called "uncertainty", to the perceived cost. This uncertainty is generally a random quantity (between days and between drivers) based on the actual variation in traffic conditions. As a result, the route flows are random variables rather than deterministic, which is why Walting (2002)'s model is considered a more general stochastic user equilibrium model.

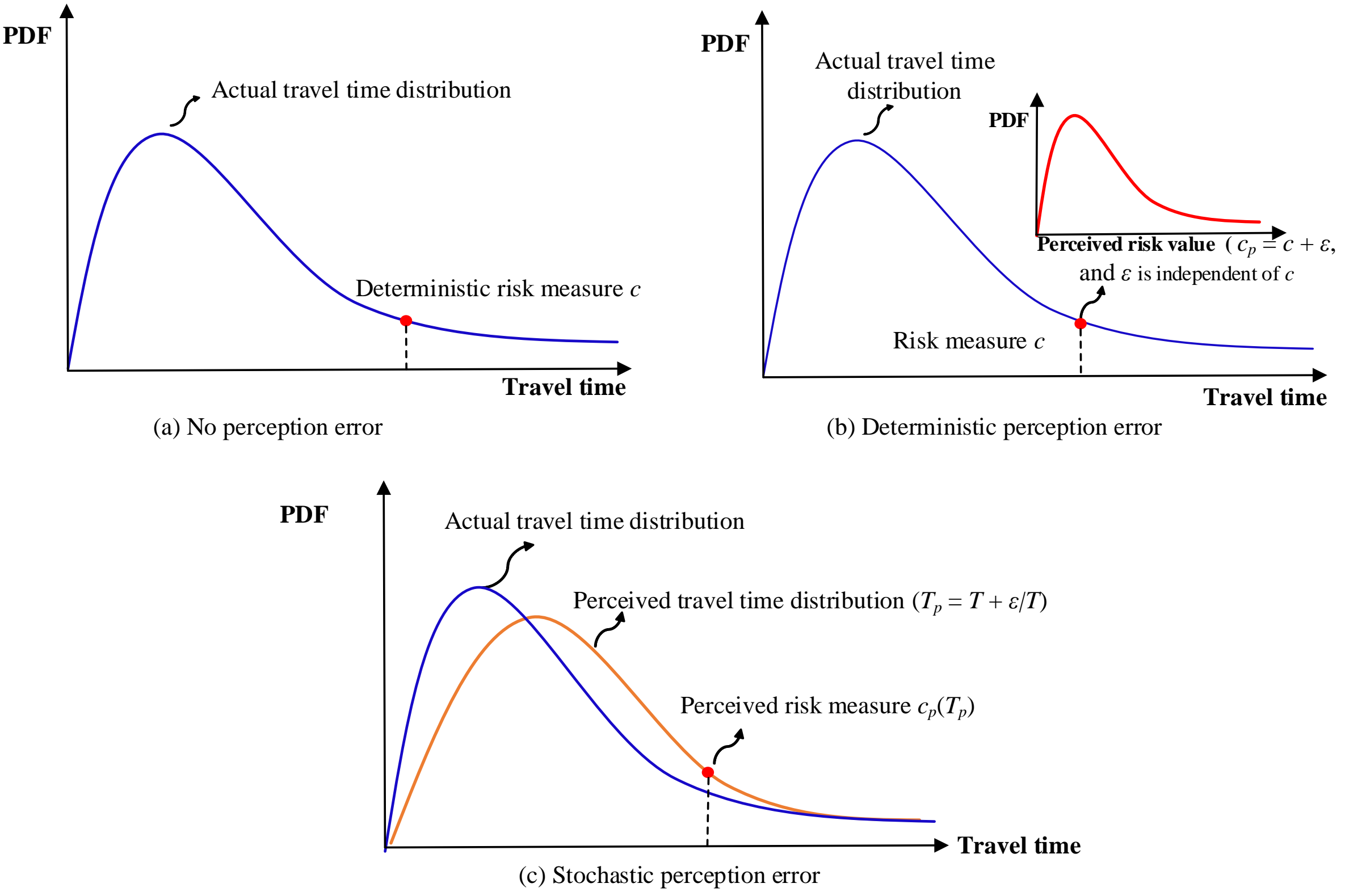

Figure 8. Three approaches to modeling travelers' perception error under travel time variability

## 5.2 Mathematical Models and Solution Algorithms

### 5.2.1 Mathematical models

Because of the non-additive route cost structure, most TTR-based traffic assignment models are formulated as a variational inequality problem (Shao *et al*., 2006a; Waltling 2006; Chen and Zhou, 2010; Chen *et al*., 2011b; Nie, 2011; Wu and Nie, 2013), a non-linear complementarity problem (Ordoñez and Stier-Moses, 2010), or a fixed point problem (Mirchandani and Soroush, 1987; Lam *et al*., 2008; Sumalee *et al*., 2011) in terms of the route flow variables. The existence of a solution is guaranteed if the route cost function is continuous and the feasible set is convex. An exception to the above route-based formulations is the link-based mean-excess traffic equilibrium model proposed by Xu *et al*. (2017). Taking advantage of the sub-additivity of the mean-excess travel time, Xu *et al*. (2017) proposes a link-based mean-excess traffic equilibrium model in the form of a Beckmann-like transformation. Below we will summarize the general formulations of the above TTR-based traffic assignment models.

*(1) Variational inequality*

The route flow vector **f*** is at equilibrium if the following variational inequality holds:

$$\mathbf{c}(\mathbf{f}^*)^{\text{Transpose}}(\mathbf{f}-\mathbf{f}^*) \geq 0,\ \forall \mathbf{f} \in \Omega_{\mathbf{f}},\ \Omega_{\mathbf{f}} = \left\{ f_{\omega j} : \sum_{j \in J_\omega} f_{\omega j} = q_\omega,\ \forall \omega \in \Omega_\omega \right\} \tag{42}$$

where $\Omega_{\mathbf{f}}$ is the feasible set of the route flow vector. Note that if different forms of route cost functions $\mathbf{c}(\cdot)$ are adopted, as shown in Section 5.1, Eq. (42) would lead to different equilibrium models in the variational inequality form.

*(2) Non-linear complementarity problem*

The route flow vector f is at equilibrium if the following non-linear complementarity problem holds: $0 \leq \mathbf{f} \perp (\mathbf{c}(\mathbf{f}) - \mathbf{c}_{\min}) \geq 0$. $\mathbf{c}_{\min}$ is the vector of the minimum travel cost for each O-D pair. Likewise, different forms of the route cost functions $\mathbf{c}(\cdot)$ would lead to different equilibrium models in the non-linear complementarity problem form.

*(3) Fixed point problem*

Traffic equilibrium problems under uncertainty, particularly stochastic user equilibrium problems, can also be formulated as fixed point problems. $\mathbf{f} = H(\mathbf{f})$ is the generic formulation of a fixed point problem, where $H(\cdot)$ is a mapping function; for instance, $H(\mathbf{f})$ may equal the product of O-D demand and route choice probability.

*(4) Game theory model*

Bell and Cassir (2002) proposes a game theory model to formulate risk-averse traffic assignment problems. The upper-level problem for the demon maximizes the total expected cost imposed on network users; while the lower-level problem for travelers is a standard deterministic user equilibrium assignment problem, where travelers select the routes with the lowest expected costs in different disruption scenarios caused by the demon. The mathematical formulations of game theory-based equilibrium model with multiple O-D pairs are as follows. For each O-D pair $\omega$, solve the following simultaneously.

$$\begin{aligned} &\text{Upper level}: \max_{\boldsymbol{\rho}_\omega} \sum_{j \in J_\omega} \sum_{s \in S_\omega} \rho_{\omega s} c_{\omega js}(\mathbf{f}) f_{\omega j},\ \text{s.t.} \sum_{s \in S_\omega} \rho_{\omega s} = 1,\ \boldsymbol{\rho}_\omega \geq 0 \\ &\text{Lower level}: \max_{\mathbf{f}_\omega} \sum_{a \in A} \sum_{s \in S_\omega} \rho_{\omega s} \int_0^{v_a(\mathbf{f})} c_{as}(x)\,dx,\ \text{s.t.}\ v_a = \sum_{j \in J_\omega} \delta_{aj} f_{\omega j},\ \sum_{j \in J_\omega} f_{\omega j} = q_\omega \end{aligned} \tag{43}$$

where $c_{\omega js}(\mathbf{f})$ and $c_{as}(x)$ are the cost of route $j$ and the flow-dependent cost on link $a$ under scenario $s$, respectively; $S_\omega$ is the set of scenarios; $\rho_{\omega s}$ is the probability of scenario $s$ (or link damage) for O-D pair $\omega$, and $\boldsymbol{\rho}_\omega$ is the corresponding vector of scenario (or link damage) probabilities for O-D pair $\omega$.

*(5) Mathematical programming*

If the route cost is additive, the equilibrium link flow pattern **v*** can be obtained by solving the following mathematical programming in the form of a Beckmann's transformation (Beckmann, 1956).

$$
\begin{aligned}
&\min_{\mathbf{v}} \sum_{a\in A} \int_0^{v_a(\mathbf{f})} t_a(x)dx \\
&\text{s.t.} \\
&v_a = \sum_{\omega\in\Omega_\omega} \sum_{j\in J_\omega} \delta_{aj} f_{\omega j}, \ \forall a \in A \\
&\sum_{j\in J_\omega} f_{\omega j} = q_\omega, \ \forall \omega \in \Omega_\omega \\
&f_{\omega j} \geq 0, \ \forall \omega \in \Omega_\omega, \ j \in J_\omega
\end{aligned}
\tag{44}
$$

Due to the non-additive cost structure (e.g., travel time budget and percentile travel time), most TTR-based traffic assignment models cannot have the above form. However, the mean-excess travel time-based equilibrium model can be approximately formulated in this way because of the sub-additivity of the mean-excess travel time (see Xu *et al*., 2017 for more details).

### 5.2.2 Solution algorithms

Table 7 summarizes the existing solution algorithms for solving TTR-based traffic assignment problems, which can be classified into two types: route-based algorithms and link-based algorithms. Because of the non-additive structure of most risk measures, the majority of studies adopt a route-based algorithm. The route-based algorithms can be further categorized into four types: the alternating direction method, the gradient projection method, the route-based method of successive averages, and optimization algorithms using gap function reformulations of the variational inequality problem. Some of the proposed approaches require enumerating routes in advance (Shao *et al*., 2006a; Chen and Zhou, 2010; Chen *et al*., 2011b), whereas others use the column generation procedures to avoid the need of a pre-specified route set (Wu and Nie, 2013; Chen *et al*., 2011c). Only a few studies use a link-based algorithm, such as the well-known Frank-Wolfe algorithm, because of additive cost structures in their models. Exceptions include Chen *et al*. (2002a), where the route utility is the exponential function of route travel time, which can be obtained by summing up the link travel time, and Xu *et al*. (2017), where the route cost is the sum of the mean-excess travel times of the links making up the route.

Although TTR-based traffic assignment problems adopt similar solution algorithms as conventional traffic assignment problems, the former are generally more computationally demanding than the latter for two reasons. First, the properties of the route cost under uncertainty may be different from the properties without uncertainty. For example, the route cost under uncertainty may not be monotonic with respect to the route flow, meaning that only a heuristic solution can be found. Second, calculating the risk measures in TTR-based traffic assignment problems requires the consideration of a series of uncertainty propagations (to be reviewed in Section 6), which is much more complex than calculating the deterministic link travel time function as used in conventional traffic assignment problems.

Table 7. Solution algorithms of TTR-based traffic assignment problems

| Type | Method | References |
|---|---|---|
| Route-based algorithm | Alternating direction | Shao *et al*. (2006b); Chen *et al*. (2010, 2011b) |
| | Gradient projection | Nie (2011); Zhang *et al*. (2011); Wu and Nie (2013); Ji *et al*. (2017) |
| | Route-based method of successive averages | Waltling (2002); Shao *et al*. (2006a, 2008) |
| | Gap function | Lo *et al*. (2006); Szeto *et al*. (2006); Siu and Lo (2008) |
| Link-based algorithm | Frank–Wolfe algorithm | Chen *et al*. (2002a); Cheu *et al*. (2007); Xu *et al*. (2017) |

### 5.3 Discussions of Travel Time Reliability-Based Traffic Assignment Models

This section briefly reviews the dynamic traffic assignment with the consideration of travel time reliability and the applications of travel time reliability-based traffic assignment model in different research problems, mainly including network design problem, road pricing, and evacuation models.

*(1) Dynamic traffic assignment with the consideration of travel time reliability*

Compared to the static traffic assignment, dynamic traffic assignment can more realistically capture some important time-dependent phenomena, especially queue formation, link spillover, and temporal bottlenecks formation, and thus has the potential to more effectively support policy evaluation and traffic operation strategy design. Specifically, travel time reliability is considered in the route choice component of the dynamic traffic assignment model. For example, Boyce *et al*. (1999) proposes a stochastic dynamic user optimal model, where it is assumed that route travel times are variable and perceived by travelers at each time instant and travelers selected routes with the minimum perceived disutilities at each time. Szeto and Sumalee (2009) and Szeto *et al*. (2011) propose the reliability-based stochastic dynamic user optimal route choice principle, which uses the effective travel time to capture travelers' attitudes towards the risk of late arrivals owing to travel time variability. Later on, Ng and Waller (2012) considers travel time reliability as the probability that the real travel time deviates from the expected value in their linear programming cell transmission -based dynamic traffic assignment model. To assess the within-day dynamics of transportation networks and travel time reliability of public railways, Nakayama *et al*. (2012) establishes a semi-dynamic traffic assignment model under stochastic travel times. Besides, Jiang *et al*. (2011) and Zockaie *et al*. (2015) develop the multi-criterion dynamic user equilibrium traffic assignment model, in which the route choice framework explicitly considers heterogeneous users who want to minimize travel time, out-of-pocket cost, and travel time reliability.

(2) *The applications of travel time reliability-based traffic assignment model*

As network users are concerned with the reliability of reaching their destinations on time, travel time reliability concept is generally embedded into the network design problem. For example, Yang *et al*. (2000) suggests that combining capacity reliability and travel time reliability together could provide a valuable tool for designing reliable transportation networks. Chootinan *et al*. (2005) formulates the reliability-based network design problem as a bi-level program of which the lower level sub-program uses the probit-based stochastic user equilibrium to capture travel time reliability. Later on, Chen *et al*. (2007) proposes alpha reliable network design model to minimize the total travel time budget required to satisfy the total travel time reliability constraint while considering network users' route choice behavior. Ng and Waller (2009) proposes a mean-variance type of system-optimal network design model with probabilistic guarantees on system-wide travel time, which could yield a confidence interval within which the total system-wide travel time lies when implementing the prescribed capacity expansion decisions. In addition, to make a tradeoff between capacity reliability and travel time reliability, Chen *et al*. (2011a) further presents a bi-objective reliable network design model to determine the optimal link capacity enhancements under travel demand uncertainty.

Besides the above network design problem, TTR-based traffic assignment has also been have incorporated into evacuation models. For instance, Waller and Ziliaskopoulos (2006) develops a system optimum-dynamic traffic assignment to provide a robust solution with a user specified level of reliability for evacuation modeling. Later on, a model proposed by Ng and Waller (2010b) could ensure a user-specified reliability level on the evacuation plan, which only requires the knowledge of the mean values and upper and lower bounds on the evacuation demand and road capacities. Ng and Lin (2015) extends their work by providing new and complementary probability inequalities for the case when only knowing the means and variances of the uncertain quantities. Lim *et al*. (2015) further considers capacity uncertainty of road links and proposes a reliability-based evacuation route planning model, which uses the breakdown minimization principle to find the reliable evacuation routes to load evacuation flow in the network.

Due to the unavoidable travel time variability, researchers also consider travel time reliability in road pricing problem. For example, Jiang *et al*. (2011) proposes a multi-criterion dynamic user equilibrium model to explicitly consider travel time reliability for designing and analyzing road pricing strategies, while Fakhrmoosavi *et al*. (2021) presents an equitable pricing scheme with total travel time as reliability measures for heterogeneous users who have different perception values of time and reliability.

## 6. Uncertainty Propagation in Modeling Travel Time Reliability

In this section, we review the methods for addressing a common difficulty in modeling TTR in transportation networks—uncertainty propagation, i.e., the propagation of uncertainty from its source to TTR at different spatial levels. Specifically, we review the uncertainty propagation from the uncertainty source to the link TTD, and subsequently to the route or network TTR.

### 6.1 Uncertainty Propagation from Source to Link TTD

In early studies on TTR-based traffic assignment, TTD is assumed to be exogenous (i.e., provided explicitly) and flow-independent. Thus, these studies do not make any assumptions about the source and there is no propagation consideration of uncertainty (Mirchandani and Soroush, 1987; Uchida and Iida, 1993). Recent studies on TTR-based traffic assignment assume that TTDs are flow-dependent and that travel time variability originates from the uncertainty in certain sources, such as the supply side (Lo and Tung, 2003; Lo *et al*., 2006), the demand side (Shao *et al*., 2006a, 2006b), or both (Lam *et al*., 2008). Figure 9 summarizes the three common paradigms for understanding the uncertainty propagation processes in TTR-based traffic assignment problems.

- In Paradigm 1, supply side uncertainty is the sole uncertainty source and includes two cases: link capacity variation (Lo and Tung 2003; Lo *et al*., 2006) and free-flow travel time variation (Chen *et al*., 2010). Under this paradigm, the uncertainty would directly propagate from the source to the link travel time through a link cost function.
- In Paradigm 2, demand side uncertainty is the sole uncertainty source (e.g., Clark and Watling, 2005; Shao *et al*., 2006a, 2006b). Under this paradigm, the uncertainty propagates first from the source to the route flow through an assumed route choice model, and then to the link travel time through the definitional constraint between route flow and link flow and the link cost function.

- In Paradigm 3, uncertainties in both the demand and supply sides are considered to be the uncertainty sources (Lam *et al*., 2008). As a result, the uncertainty propagation process in this paradigm combines those of the above two paradigms. Interested readers may refer to Lam *et al*. (2008) for more details.

Note that the general formulae for characterizing the TTR based on the above uncertainty sources have been provided in Sections 2.2.4 and 3.2.1.

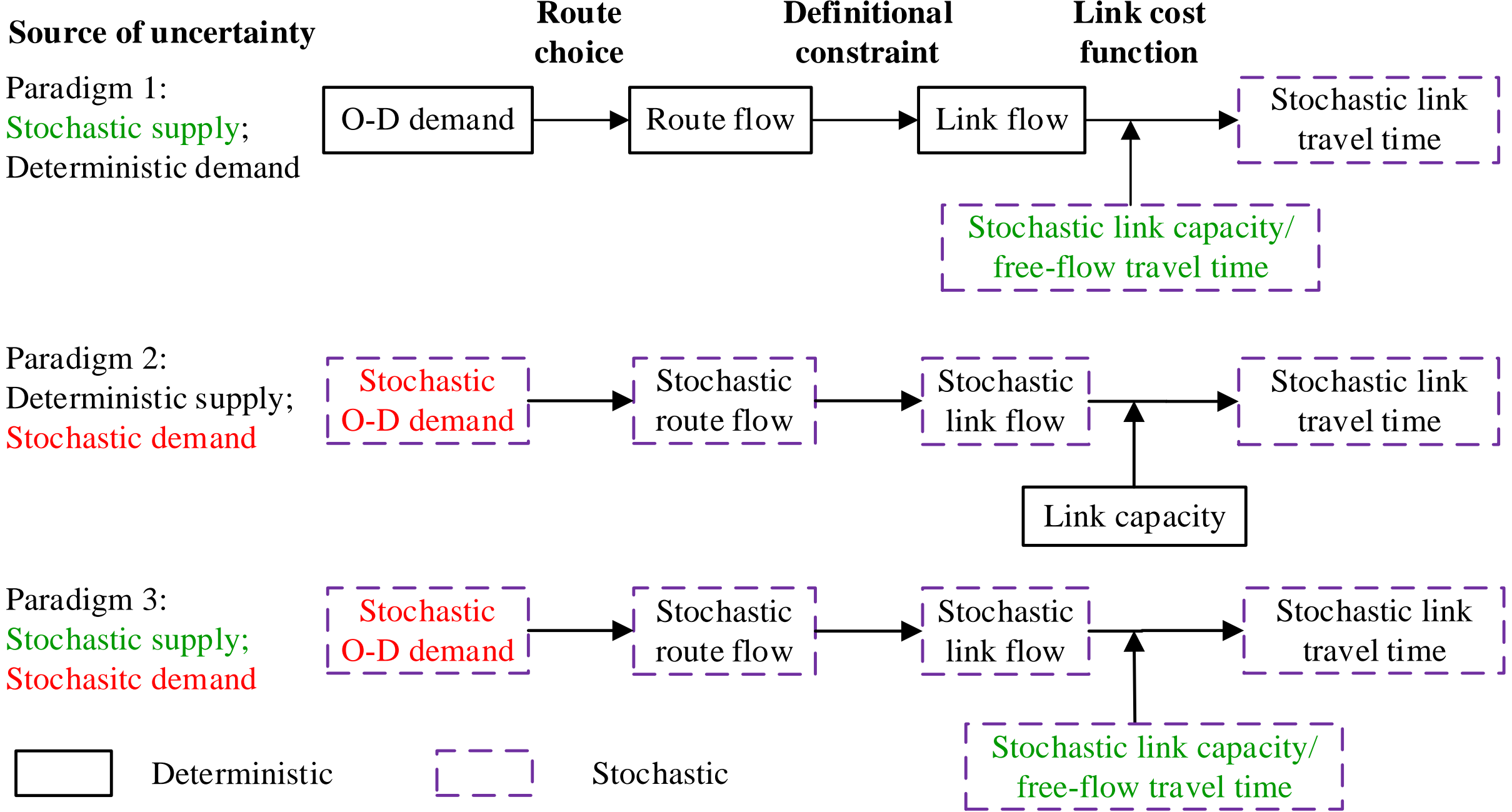

Figure 9. Three paradigms for understanding uncertainty propagation in traffic assignment under travel time uncertainty

## 6.2 Uncertainty Propagation from Link TTD to Route (Network) TTR

Generally, there are two ways of modeling uncertainty propagation from link TTD to route (network) TTR, as summarized in Figure 10. The first way shown by the dashed line in Figure 10 avoids TTD aggregation from link to route/network and directly obtains the route or network TTR from the link TTR. The second way showed by the solid line in Figure 10 considers the aggregation of TTD from the link to route (or network) level, and then the route or network TTR is obtained via the corresponding TTD. The methods for capturing uncertainty propagation under these two ways are summarized in Table 8, indicating their inputs and outputs, spatial and temporal dependence, TTD assumptions, references and notes.

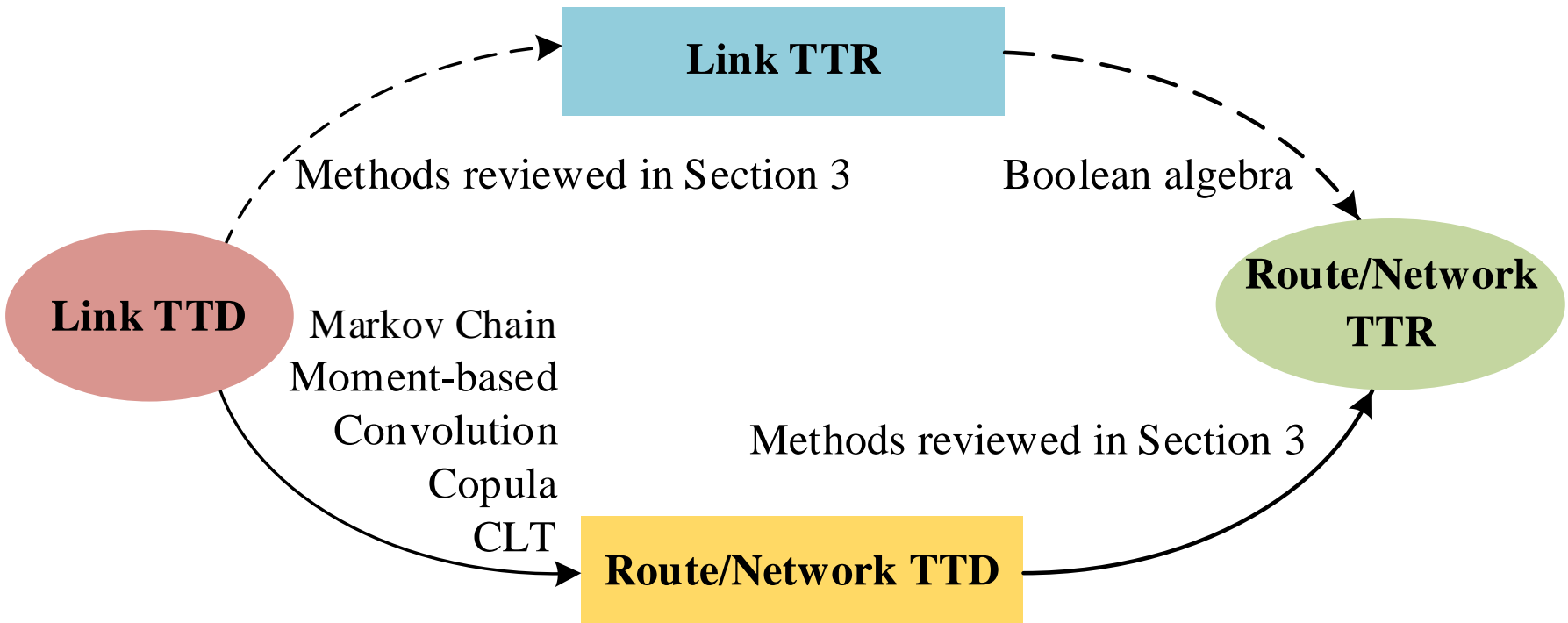

Figure 10. Typical ways of aggregating the uncertainty from link TTD to route/network TTR

### 6.2.1 From Link TTD to Route (Network) TTD and to Route (Network) TTR

Aggregating TTD from link to route/network requires a trade-off between computational efficiency and modeling accuracy, and appropriately handling the correlations between link TTDs is a key to this process. Generally, studies handle the correlations in three ways: (1) neglecting the correlations between link TTDs; (2) considering the correlations between link TTDs with TTD assumptions; and (3) considering the correlations between link TTDs without TTD assumptions.

*(1) Neglecting the correlations between link TTDs*

These methods sacrifice modeling flexibility/realism for mathematical tractability. For example, many studies adopt the Central Limit Theorem when aggregating link TTDs to route TTDs (Lo and Tung, 2003; Lo *et al*., 2006; Shao *et al*., 2006; Chen and Zhou, 2010), especially in TTR-based traffic assignment problems. If link TTDs are independently and identically distributed (i.i.d.), the Central Limit Theorem would make the route TTDs follow the Normal distribution. Namely, we have

$$T_j = \sum_{a\in A} T_a \delta_{aj} \sim N\left( \sum_{a\in A} \mu_a \delta_{aj},\ \sum_{a\in A} \sigma_a^2 \delta_{aj} \right) \tag{45}$$

Some variants of the Central Limit Theorem weaken the strict independently and identically distributed assumption. For example, link TTDs have to be independent but not necessarily identically distributed in the Lyapunov Central Limit Theorem and Lindeberg Central Limit Theorem (Nie, 2011). Although these variants to some extent weaken the independently and identically distributed assumption, empirical datasets may still not satisfy the remaining assumptions. Xu *et al*. (2017) shows that the route TTD does not always follow a Normal distribution, particularly when the number of links on a route is less than 30. This means that the Normal distribution assumption of route TTD in the Central Limit Theorem may sacrifice too much accuracy for computational efficiency, and thus it fails to capture the empirical characteristics of TTDs (i.e., positive skew and long upper tail). Furthermore, link TTDs are generally interdependent in reality, while the Central Limit Theorem requires them to be independent. This inconsistency also leads to a loss of accuracy in the aggregation process.

Another intuitive method for aggregating link TTD is the convolution integral (Fan *et al*., 2005; Ng *et al*., 2010; Nie, 2011; Ramezani and Geroliminis, 2012; Yang *et al*., 2013), which assumes that link TTDs are independent. Compared to the Central Limit Theorem, the convolution integral has two advantages: (1) it does not require all of the link TTDs to be identically distributed, and (2) it does not have a requirement on the number of links to ensure its accuracy. Considering that a route $j$ consists of $n$ links, then the PDF of the route TTD is

$$f_j(T) = f_1(T) * f_2(T) * \cdots * f_n(T), \text{ where } f_i(T) * f_{i+1}(T) \triangleq \int_{-\infty}^{\infty} f_i(x) f_{i+1}(T-x)\,dx \tag{46}$$

where $f_i(\cdot) * f_{i+1}(\cdot)$ represents the convolution integral of two link TTDs. It is time-consuming when routes consist of more than two links because of the recursive integral, as revealed in Eq. (46). Therefore, some researchers use approximation algorithms to reduce the computational burden, such as the Laplace transform (Fan *et al*., 2005) and the fast Fourier transform (Ng *et al*., 2010). However, link TTDs in real transportation networks are usually interdependent, and the convolution integral ignores the spatiotemporal correlations between link TTDs, which may lead to errors in obtaining route TTDs (Ramezani and Geroliminis, 2012; Chen *et al*., 2017).

(2) *Considering the correlations between link TTDs with TTD assumptions*

To alleviate the loss of accuracy caused by neglecting the correlations between link TTDs, some researchers concentrate on modeling the correlations in TTD aggregation with assumed distribution type of link travel time.

Lognormal distribution is a typically assumed distribution type for link TTD for capturing the correlations between link TTDs. Then, the Fenton–Wilkinson approximation is used for obtaining route TTD (Srinivasan *et al*., 2014; Chen *et al*., 2018). This approximation method is first proposed by Fenton (1960) to model the transmission loss in communication engineering. Fenton (1960) numerically finds that the sum of several Lognormal random variables is still a Lognormal distribution. That is, the mean of the route TTD equals the sum of the mean of the component link TTDs, and the variance equals the sum of all of the elements of the covariance matrix of the component link TTDs. Based on the Fenton–Wilkinson approximation, the parameters of route TTD can be obtained from the parameters of the link TTDs. Modeling flexibility has been further expanded through the use of 3-parameter

Lognormal (i.e., shifted Lognormal) distributions (Srinivasan *et al*., 2014). This new shift parameter of route TTD can also be expressed by the shift parameters of the link TTDs.

In a nutshell, the Fenton–Wilkinson approximation uses the approximated additivity of a Lognormal distribution. Similarly, Castillo *et al*. (2013) uses the property of location-scale family distributions and assumes that the route TTD follows a location-scale family distribution whose parameters can be obtained with the mean and covariance of the link TTDs. The Fenton–Wilkinson approximation is a special case of the moment generating function (MGF) method that is often used to compute a distribution's moments (Rudin, 1976). The MGF is the expectation of a function of the random variable, which can be defined as follows:

$$MGF_T\left(x\right) = E\left[e^{xT}\right] \tag{47}$$

MGF is an alternative expression of a distribution that uniquely determines the distribution. Ma *et al*. (2017) adopts the MGF method for approximating the sum of the correlated Normal and Lognormal link TTDs. The route TTD can be derived using its MGF, and its MGF can be calculated from the component link TTDs: $\mathrm{MGF}_{\mathrm{route}} = MGF_T\{T_1, T_2, \ldots, T_{\mathrm{n}}\}$.

Note that MGF may not have a general closed form; in such cases, MGF can be approximated by other methods. For example, the MGF of a Lognormal distribution does not have a closed form, so Mehta *et al*. (2007) uses a Gauss–Hermite expansion to approximate it. However, not all distributions have MGFs.

*(3) Considering the correlations between link TTDs without TTD assumptions*

As a powerful tool to describe the dependency between random variables, the copula function is a more general method for modeling the aggregation of link TTDs (e.g., Chen *et al*., 2017; Chen, *et al*., 2019; Luan *et al*., 2019; Yun *et al*., 2019; Prokhorchuk *et al*., 2019; Qin *et al*., 2020; Samara *et al*., 2020; Yu *et al*., 2020). Before being adopted in transportation research, the copula function has been widely used in finance and insurance (Nelsen, 2006). According to Sklar's theorem, a copula function can be used to connect multiple univariate marginal distributions to express a corresponding multivariate joint distribution (Nelsen, 2006). Following this rule, the joint TTD of a route consisting of *n* links can be expressed by marginal link TTDs with the help of copula function *C*:

Table 8. Methods for determining uncertainty propagation from the link level to the route (network) level

| Method | Input | Output | Dependence | | TTD assumption | Literature | Note |
|---|---|---|---|---|---|---|---|
| | | | Spatial | Temporal | | | |
| Central limit theorem (CLT) | Mean and variance of link TTD | Route/network TTD | No | No | Link TTD i.i.d.; Route TTD follows a normal distribution | Lo and Tung (2003); Lo *et al*. (2006); Shao *et al*. (2006a, 2006b); Chen and Zhou (2010) | In some variants of CLT, such as Lyapunov CLT and Lindeberg CLT, link TTDs are not required to be identically distributed |
| Convolution integral | Link TTD | Route TTD | No | No | Link TTDs are independent | Fan *et al*. (2005); Ng *et al*. (2010); Nie (2011); Ramezani and Geroliminis (2012); Yang *et al*. (2013); Filipovska and Mahmassani (2020); Filipovska et al. (2021) | Different algorithms are used to calculate convolution integral, e.g., the Laplace Transform and the Fast Fourier Transform |
| Moment-Based | Moments and covariance matrix of link travel time | Route/network TTD | Yes | No | Link TTD follows a multivariate normal or lognormal distribution | Srinivasan *et al*. (2014); Ma *et al*. (2017); Chen *et al*. (2018); Chen *et al*. (2020) | Some also assume that route TTD follows a lognormal distribution |
| Copula function | Link TTD and copula function | Route TTD | Yes | Yes | No | Chen *et al*. (2017); Chen, *et al*. (2019); Luan *et al*. (2019); Yun *et al*. (2019); Prokhorchuk *et al*. (2019); Qin *et al*. (2020); Samara *et al*. (2020); Yu *et al*. (2020) | / |

| Markov Chain framework | Traffic states; transition probability; and link TTD | Route TTD | Yes | Yes | No | Ramezani and Geroliminis (2012); Ma *et al*. (2017); Yu *et al*. (2020) | Often used in combination with other aggregation methods |
|---|---|---|---|---|---|---|---|
| Empirical evidence | Mean and variance of link TTD | Mean and variance of route TTD | Yes | No | No | Rakha *et al*. (2006); Kaparias *et al*. (2008) | Route coefficient of variation is the mean coefficient of variation over all links |
| Boolean algebra | Link reliability function | Network reliability | No | No | Independent/ dependent | Al-Deek and Emam (2006); Emam and Al-Deek (2006) | Link reliability function can be expressed by using the well-defined reliability engineering functions |

Note. i.i.d : independently and identically distributed.

$$f(T_1, T_2, \cdots, T_n; \theta) = C\left(F(T_1, \theta_1), F(T_2, \theta_2), \cdots, F(T_n, \theta_n); \theta_c\right)\prod_{i=1}^{n} f(T_i; \theta_i) \tag{48}$$

where $f(T_1, T_2, \cdots, T_n; \theta)$ represents the joint route PDF with the parameter $\theta$; $f(T_i, \theta_i)$ and $F(T_i, \theta_i)$ are the marginal link PDF and CDF with the parameter $\theta_i$, respectively. Many copula families have been developed in the literature and interested readers may refer to Bhat and Eluru (2009) for bivariate copulas and Luan *et al*. (2019) for multivariate copulas. Generally, modeling the aggregation of link TTDs using copula functions requires three steps: (1) measuring the dependency between link TTDs, (2) selecting optimal copula functions, and (3) modeling the route TTD. The computation used in the copula-based method is challenging in higher dimensions (i.e., routes consisting of more than two links) and therefore is cumbersome for large-scale networks. Yun *et al*. (2019) proposes a pair-copula to overcome this challenge, and further verification of its performance with a larger number of links is needed.

Another method used in modeling TTD aggregation is the Markov chain (Ramezani and Geroliminis, 2012; Ma *et al*., 2017; Yu *et al*., 2020), which is used to describe the stochastic process where one state transitions to another based on corresponding probabilistic rules. The method using Markov chain consists of three stages: (1) defining traffic states, (2) estimating the transition probability of traffic states, and (3) estimating the route TTD. Generally, a Markov chain serves as the framework for the whole TTD aggregation, and it is often combined with other aggregation methods in stage three to estimate Markov route TTDs. For instance, Ramezani and Geroliminis (2012), Ma *et al*. (2017), and Yu *et al*. (2020) respectively adopt convolution integral, MGF, and copula in stage three of the Markov chain framework to aggregate link TTDs.

Note that this section only reviews methods that explicitly consider link TTD aggregation. Some studies propose methods to fit route/trip TTD directly based on travel time data, which may implicitly involve link TTD aggregation, e.g., the generative adversarial network method to model trip TTD with GPS data in Zhang *et al*. (2019a), and the statistical model to characterize route TTD for probe vehicle data in Jenelius and Koutsopoulos (2013).

### 6.2.2 From Link TTD to Link TTR and to Route (Network) TTR

As shown in Figure 10, directly obtaining route or network TTR from link TTR is another approach to aggregating TTR that can circumvent calculating/fitting route or network TTDs. These methods require that link TTR is well-defined in terms of additivity or other similar properties. However, few studies adopt such methods as it is difficult to satisfy these criteria mathematically.

A typical method is to use the variance as the reliability measure and then it is straightforward to calculate the route TTR from the link TTR because the variance is additive (Sen *et al*., 2001; Yin and Hitoshi, 2001). To circumvent storing the covariance matrix in estimating the variance of route travel time (Rakha *et al*., 2006; Kaparias *et al*., 2008), some studies use empirical approximations which assumes that the coefficient of variation of the route TTD is the mean coefficient of variation of all of the component links.

Another method is to calculate the route or network TTR based on the probability of failure or hazard of links in terms of travel time from the perspective of system reliability engineering. In fact, this method is widely used to assess connectivity reliability (Bell and Iida, 1997). Specifically, the calculation uses Boolean algebra based on the configuration of the networks (i.e., series configuration, parallel configuration, or both) (Al-Deek and Emam, 2006; Emam and Al-Deek, 2006). First, the state of link $a$ can be defined as a binary variable based on link TTR: $R_a = 1$ if link $a$ functions and $R_a = 0$ otherwise. Then, the route TTR $R_j$ with series and parallel configurations can be expressed by the following equation:

$$R_j = \begin{cases} \prod_{a \in A} R_a & \text{for series configuration} \\ 1 - \prod_{a \in A} \left(1 - R_a\right) & \text{for parallel configuration} \end{cases} \tag{49}$$

The computational burden of Boolean algebra would be high if the network consists of complicated combinations of series and parallel configurations.

## 7. Conclusions and Future Research

This paper reviewed the methodological developments of modeling TTR in transportation networks. Adopting a network perspective, this review concentrated on the methods of modeling TTR from depicting the whole variability picture (i.e., characterizing TTDs) to assessing the TTR (i.e., TTR evaluation and TTR valuation) and lastly to investigating the effects of TTR on individual users' travel behavior and the collective network flow pattern (i.e., TTR-based traffic assignment). Also, this paper reviewed the methods for addressing a common challenge for modeling TTR in transportation networks: uncertainty propagation from the uncertainty source to link TTR and to route/network TTR. This review, particularly with the integrated review framework and network-wide perspective, is expected to provide a better and deeper understanding about the relationships and common components of four topics mainly studied from different disciplines, and to help developing more possible combinations of TTR modeling philosophy.

Specifically, this paper summarized two common modeling frameworks for characterizing TTD (i.e., fitting and deducing the TTD) and three difficulties in applying these frameworks: (1) characterizing heterogeneous TTDs, 2) satisfying different requirements for theoretical and empirical applications, and (3) deriving route or network TTD based on link TTD. Under these two frameworks, this paper identified four modeling rationales and provided their general formulas and corresponding TTD models. To make the abstract characterizations of TTD understandable and intuitive, TTR evaluation quantitatively assesses the reliability performance using various reliability measures, and TTR valuation focuses on quantifying the VOR in monetary units to understand users' responses. In particular, TTR evaluation can be directly based on empirical datasets or on assumed uncertainty sources. Therefore, we presented the general formulas of the methods under these two perspectives and summarized the associated behavior assumption, consistency, and accuracy of the reliability measures. In summarizing the valuation of TTR, this paper reviewed the four mathematical models used to measure VOR: the mean-variance model, schedule delay model, mean-lateness model, and network utility maximization model. Then, we summarized the valuation measures and presented the dimensions of the VOR according to the general theoretical frameworks used to derive the valuations. To further investigate the effects of TTR on travelers' individual route choice behaviors and collective network flow pattern, we reviewed the route choice criteria and corresponding TTR-based assignment models as well as their solution algorithms. Two key dimensions in modeling travelers' route choice criteria are identified, i.e., objective travel time variability and subjective perception error.

Although this paper reviewed a number of novel ideas and approaches for modeling TTR in transportation networks, many areas remain open to be further investigated. Some potential directions for future research close to the reviewed research topics of this paper are as follows.

- *Unified approach with closed-form expressions for characterizing TTDs in transportation networks*. Although many different TTD models based on different modeling rationales have been developed, it is hard to identify an optimal distribution function for heterogeneous TTDs. In addition, conceptualizing and formulating a computationally tractable or analytically derived TTR model needs closed-form expressions of TTD models. Therefore, it is necessary to develop a unified approach with closed-form expressions to characterize heterogeneous TTDs in transportation networks. To this end, Zang *et al.* (2018b) makes some attempts but the proposed model only has closed-form expressions of PPF. More efforts in this direction are needed for exploring other possible ways and promoting TTR applications in large-scale networks.

- *Simple but useful criteria to select reliability measures*. As discussed in Section 3.3, the existing reliability measures may not behave consistently for the same assessment object. Therefore, it is necessary to establish some criteria to support selecting or developing an optimal reliability measure for practical applications. These criteria should be simple to be well understood by users but useful to identify different optimal reliability measures for different application purposes. To unify and streamline the pool of TTR measures, we need to have a complete and quality dataset to explore the relationship and connection among the existing measures.
- *Reasonable network-wide TTR monitoring method*. In the literature, when analyzing the network-wide TTR, we usually compute some reliability measures with respect to total travel time. Total travel time is the product of link flow and link travel time, which is a simple aggregation from link level to network level. However, this is not suitable for network-wide TTR monitoring. On the one hand, this measure is too aggregated due to the summation and multiplication, making it insensitive to identify network perturbations. On the other hand, it cannot easily tell the source of an identified abnormal network state. Hence, innovative methods are needed to achieve this purpose.
- *Quantifying the VOR at route or network level*. Although the value of travel time (VOT) and VOR are critical to many traffic network models, there are very limited studies to estimate VOT and VOR at network level. Accordingly, the VOT and VOR are generally given and assumed values in network models. Uchida (2014) firstly estimates the VOT and VOR in transportation networks considering the network structure and drivers' route choice behavior. However, how to quantify the VOR at route, O-D, or network level remains to be open for further investigation. Furthermore, it is important to examine the relationship between classical valuation models at trip level and the new network-level valuation models.

The following identifies some potential future research directions that involve more topics of modeling TTR in transportation networks, especially in the era of new data environment, applications, and emerging technologies.

- *Constructing large-scale baseline travel time datasets at multiple spatial levels*. Due to the lack of large-scale datasets of transportation networks, it is hard to testify the validation of the TTR models developed based on various assumptions and also hard to conduct fair comparisons of TTR models. The advent of big data is making large-scale travel time datasets at the route/network level available, therefore it is necessary to construct the baseline travel time datasets for the development of modeling TTR in transportation networks. The datasets should be well customized to suit different application purposes, such as TTD characterization at heterogenous links, routes and time periods, and route choice modeling under uncertainty. Based on the datasets, several unique and less studied research questions could be investigated, e.g., (1) the optimal selection of time aggregation interval of modeling TTR for different application purposes, (2) validating the modeling of uncertainty propagation at different spatial levels by "matching" travel time data and traffic flow data; (3) studying the feasibility of still using the deterministic link travel time function like BPR and developing possible adjustment schemes of BPR parameters; and (4) developing data and model integrative driven methodologies for estimating stochastic O-D demand, route flow and travel time. The large sample size of big data is closely related to stochasticity, which should not be simply averaged and then input into the traditional deterministic models.

- *Estimating the values of the reliability/risk parameters through a non-answered way in the era of big data*. This review shows that many TTR models need travelers' risk parameters as inputs, such as the preference parameters in VOR models, the probability required in reliability measures, and the risk parameters in route choice models. However, it is difficult to set these values in advance and it is impossible to collect each user's parameters via surveys. Travelers even may not know their own exact values and these parameters may be flow-dependent rather than fixed. The data acquisition techniques, such as GPS, AVI, RFID, Bluetooth, cellular data, can automatically collect long-term massive individual datasets. It will be very promising to study how to mine (e.g., via deep learning) the travelers' reliability/risk parameters through a non-answered way.
- *Modeling the TTR of multimodal transportation networks*. Most studies of TTR concentrate on a single travel mode, but with the fast development of city scale and urban agglomeration, many trips involve more than one travel mode. Also, emerging trip-oriented services, such as mobility as a services (MaaS) and route guidance, highlight the need to understand the reliability performance of multimodal urban/regional/international transportation systems. Under a multimodal transportation system, travelers may not be able to complete their whole trip if they cannot access one of the necessary travel modes, and thus it would be important to develop methods for modeling reliability on routes involving multiple travel modes, which could be called "door-to-door TTR". Assessing a door-to-door TTR that consists of different travel modes (e.g., car, bus, metro, bicycle, shared mobility, train, air traffic, etc.) requires redeveloping the methods of assessing TTR with both flexible and scheduled services and modeling the uncertainty propagation reviewed in this paper. From the viewpoint of service providers, it is also interesting to design the insurance packages of on-time arrival of MaaS and on-time delivery of app-based services, given that the travel time is highly uncertain.

- *Investigating the challenges and benefits of emerging technologies on modeling TTR*. The popularization of emerging technologies, such as connected and automatic vehicles, can reduce traffic congestion and road crashes (Poczter and Jankovic, 2014; Fagnant and Kockelman, 2015; Papadoulis *et al*., 2019), leading to less supply-side uncertainty than in systems for conventional vehicles. However, before achieving the highest level of automation (i.e., full self-driving), the road traffic will still be mixed with both conventional vehicles and connected and automatic vehicles. During this long transition period, mixed traffic may generate new driving risk. As traffic incidents contribute significantly to the supply-side uncertainty of transportation networks (van Lint *et al*., 2008; Chen and Zhou, 2010), it would be worthwhile to explore how the emerging technologies at different transition stages could affect the TTR of transportation networks.

## SUMMARY OF ACRONYMS AND NOTATIONS AND DEFINITIONS

Table A9. Summary of acronyms.

| Acronyms | Note | Acronyms | Note |
|---|---|---|---|
| CDF | cumulative distribution function | TTR | Travel time reliability |
| DTA | dynamic traffic assignment | TTD | Travel time distribution |
| MGF | moment generating function | O-D | origin-destination |
| PDF | probability density function | UE | user equilibrium |
| PPF | percentile point function | VOR | Value of travel time reliability |
| SUE | stochastic user equilibrium | VOT | Value of travel time |

Table A2. Summary of notations and definitions.

| Notation | Definition |
|---|---|
| $z$ | The general formula of the PDF of travel time distribution |
| $T$, $\overline{T}$ | Random travel time, and the specified travel time or threshold |
| $t$ | Time |
| $f(T)$, $F(T)$, $F^{-1}(\rho)$ | PDF, CDF and inverse CDF of *travel time T* |
| $\varphi(X)$, $\Phi(X)$, $\Phi^{-1}(\rho)$ | PDF, CDF and inverse CDF of standardized *travel time X* |
| $\mu$, $\sigma$, $\sigma^2$ | Mean, standard deviation, and variance |
| $l$, $\theta$, $k$ | Location factor, scaling factor, and shape factor |
| $\xi$ | Moment of travel time |
| $\lambda$ | Weight/degree |
| $\rho$ | Probability or confidence level or reliability requirement |
| $a$, $A$ | Link and the set of links |

| $t_a(\cdot)$, $v_a$, $C_a$, | Link travel time function, link volume, and link capacity of link $a$ |
|---|---|
| $f$, $q$ | Path flow and O-D demand |
| $e$, $Q$ | Excess demand and a constant larger than the maximal traffic demand |
| ~ | Stochasticity of a variable due to assumed source of uncertainty |
| $i$, $n$ | Count number and constant number |
| $\Omega$ | Set |
| $\omega$, $\Omega_\omega$ | O-D pair and the set of O-D pairs |
| $j$, $J_\omega$ | Path and the set of paths of O-D pair $\omega$ |
| $E$, $Var$ | Expectation and variance operators |
| $\delta_{aj}$ | Link-path incidence indicator |
| $N$ | Normal distribution |
| $R$ | General formulation of travel time reliability |
| $\delta$ | Additional travel time |
| $\alpha$, $\beta$, $\gamma$ | Scheduling preference parameters in the schedule delay model |
| $\eta$ ($\eta_1$-$\eta_2$-$\eta_3$-$\eta_4$) | Preference parameter in valuation (utility) models |
| $b(\rho)$, $\eta(\rho)$ | TTB and METT for a desired reliability requirement $\rho$ |
| $U$, $EU$, $U_{cd}$ | Utility, expected utility and Cobb–Douglas utility |
| $c$ | travel cost (function) |
| $Arr$, $D$, $D^*$ | Arrival time, departure time and optimal departure time |
| $h(t)$, $w(t)$ | Utility of staying at home and staying at work at time $t$ |
| $H$, $L^+$ | Mean lateness factor, and lateness at boarding |
| $\varphi_1$-$\varphi_2$-$\varphi_3$ | Lagrange multiplier |
| $\mu_T$, $\sigma_T$ | Valuation measures for VOT and VOR |
| $\varepsilon$ | perception error |
| $H(\cdot)$ | Mapping function in the fixed point problem |
| * | Convolution operator |
| $s$ | State/scenario |
| $MGF\{\cdot\}$ | Moment generating function |
| $C(\cdot)$ | Copula function |

**ACKNOWLEDGMENT**

The authors are grateful to the four anonymous referees for their comments and suggestions to improve the paper. This study was sponsored by the National Natural Science Foundation of China (72021002) and the Shanghai Rising-Star Program (20QA1409800). These supports are gratefully acknowledged.